\documentclass[12pt]{iopart}
\usepackage{color}
\usepackage{graphicx}
\usepackage{epsf}
\usepackage{graphicx,epsfig}
 \def\neq {\not\equiv}

\def\cs2{c_{s}^{2}}

 \def\al{\alpha}
 \def\b{\beta}
 
 \def\de{\delta}
 
 \def\ep{\varepsilon}

 \def\df{\delta\phi}
 \def\p{\partial}

 \def\epr{\eta^{'}}
 \def\eps{\eta^{''}}

\def\DB{\delta B}
 \def\be   {\begin{equation}}   \def\ee   {\end{equation}}
 \def\ba   {\begin{array}}      \def\ea   {\end{array}}
 \def\bea  {\begin{eqnarray}}   \def\eea  {\end{eqnarray}}
 \def\bean {\begin{eqnarray*}}  \def\eean {\end{eqnarray*}}
\begin{document}

\title{Non-Gaussianity and statistical anisotropy from vector field populated inflationary models}

\vspace{0.8cm}

\author{ Emanuela Dimastrogiovanni$^{1,2}$, Nicola Bartolo$^{1,2}$, Sabino Matarrese$^{1,2}$ and 
Antonio Riotto$^{2,3}$}
\vspace{0.4cm}
\address{$^1$ Dipartimento di Fisica ``G. Galilei'', Universit\`{a} degli Studi di 
Padova, \\ via Marzolo 8, I-35131 Padova, Italy} 
\address{$^2$ INFN, Sezione di Padova, via Marzolo 8, I-35131 Padova, Italy}
\address{$^3$ CERN, Theory Division, CH-1211 Geneva 23, Switzerland}
\eads{\mailto{nicola.bartolo@pd.infn.it}, \mailto{dimastro@pd.infn.it}, 
\mailto{sabino.matarrese@pd.infn.it} and \mailto{riotto@mail.cern.ch}}

\date{\today}
\vspace{1cm}
\begin{abstract} We present a review of vector field models of inflation and, in particular, of the statistical anisotropy and non-Gaussianity predictions of models with $SU(2)$ vector multiplets. Non-Abelian gauge groups introduce a richer amount of predictions compared to the Abelian ones, mostly because of the presence of vector fields self-interactions. Primordial vector fields can violate isotropy leaving their imprint in the comoving curvature fluctuations $\zeta$ at late times. We provide the analytic expressions of the correlation functions of $\zeta$ up to fourth order and an analysis of their amplitudes and shapes. The statistical anisotropy signatures expected in these models are important and, potentially, the anisotropic contributions to the bispectrum and the trispectrum can overcome the isotropic parts.\\

DFPD 2010-A-01 / CERN-PH-TH 2010-010

\end{abstract}

\maketitle

\section{Introduction}

In the standard cosmological model, at very early times the Universe undergoes a quasi de Sitter exponential expansion driven by a scalar field, the inflaton, with an almost flat potential. The quantum fluctuations of this field are thought to be at the origin of both the Large Scale Structures and the Cosmic Microwave Background (CMB) fluctuations that we are able to observe at the present epoch \cite{lrreview}. CMB measurements indicate that the primordial density fluctuations are of order $10^{-5}$, have an almost scale-invariant power spectrum and are fairly consistent with Gaussianity and statistical isotropy \cite{smoot92,bennett96,gorski96,wmap3,wmap5}. All of these features find a convincing explanation within the inflationary paradigm. Nevertheless, deviations from the basic single-(scalar)field slow-roll model of inflation are allowed by the experimental data. On one hand, it is then important to search for observational signatures that can help discriminate among all the possible scenarios; on the other hand, it is important to understand what the theoretical predictions are in this respect for the different models. \\

\noindent Non-Gaussianity and statistical anisotropy are two powerful signatures. A random field is defined ``Gaussian'' if it is entirely described by its two-point function, higher order connected correlators being equal to zero. Primordial non-Gaussianity \cite{review,prop} is theoretically predicted by inflation: it arises from the interactions of the inflaton with gravity and from self-interactions. However, it is observably too small in the single-field slow-roll scenario \cite{Acqua,Maldacena:2002vr,Seery:2006vu}. Alternatives to the latter have been proposed that predict higher levels of non-Gaussianity such as multifield scenarios \cite{Linde:1985yf,Kofman:1985zx,Polarski:1994rz,GarciaBellido:1995qq,Mukhanov:1997fw,Langlois:1999dw,Gordon:2000hv}, curvaton models \cite{Mollerach,Enqvist:2001zp,Lyth:2001nq,Lyth:2002my,Moroi:2001ct,Bartolo:2003jx} and models with non-canonical  Lagrangians \cite{Garriga:1999vw,ArmendarizPicon:1999rj,Ali,Chen:2006nt,ghostinfl}. Many efforts have been directed to the study of higher order (three and four-point) cosmological correlators in these models \cite{Bartolomulti,VernizziWands,ChenWang,Barttrisp,Chen:2006nt,Seery:2006vu,SL3,Byrnes,Seery:2008ax,SVW,Huang:2006eha,Arroja:2008ga,Arroja:2009pd,Chen:2009bc,Gao:2009gd,Mizuno:2009mv,Okamoto:2002ik,Kogo:2006kh} and towards improving the prediction for the two-point function, through quantum loop calculations \cite{Schwinger:1960qe,Calzetta:1986ey,Jordan:1986ug,Maldacena:2002vr,Weinberg:2005vy,Weinberg:2006ac,Seery:2007we,Seery:2007wf,Dimastrogiovanni:2008af}. From WMAP, the bounds on the bispectrum amplitude are given by $-4<f_{NL}^{loc}<80$ \cite{Smith:2009jr} and by $-125<f_{NL}^{equil}<435$ \cite{Senatore:2009gt} at $95\%$ CL, respectively in the local and in the equilateral configurations. For the trispectrum, WMAP provides $-5.6 \times 10^5<g_{NL}<6.4 \times 10^5$ \cite{Vielva:2009jz} ($g_{NL}$ is the ``local'' trispectrum amplitude from cubic contributions), whereas from Large-Scale-Structures data $-3.5\times 10^5<g_{NL}<8.2\times 10^5$ \cite{Desjacques:2009jb}, at $95\%$ CL. Planck \cite{http://planck.esa.int/} and future experiments are expected to set further bounds on primordial non-Gaussianity.\\
     
\noindent Statistical isotropy has always been considered one of the key features of the CMB fluctuations. The appearance of some ``anomalies'' \cite{de OliveiraCosta:2003pu,Vielva:2003et,Eriksen:2003db} in the observations though, after numerous and careful data analysis, suggests a possible a breaking of this symmetry that might have occurred at some point of the Universe history, possibly at very early times. This encouraged a series of attempts to model this event, preferably by incorporating it in theories of inflation. Let us shortly describe the above mentioned ``anomalies''. First of all, the large scale CMB quadrupole appears to be ``too low'' and the octupole ``too planar''; in addition to that, there seems to exist a preferred direction along which quadrupole and octupole are aligned \cite{Bennett:1996ce,Spergel:2003cb,de OliveiraCosta:2003pu,Efstathiou:2003tv,Land:2005ad}. Also, a ``cold spot'', i.e. a region of suppressed power, has been observed in the southern Galactic sky \cite{Vielva:2003et,Cruz:2006fy}. Finally, an indication of asymmetry in the large-scale power spectrum and in higher-order correlation functions between the northern and the southern ecliptic hemispheres was found \cite{Hansen:2004mj,Eriksen:2003db,Hansen:2004vq}. Possible explanations for these anomalies have been suggested such as improper foreground subtraction, WMAP systematics, statistical flukes; the possibilities of topological or cosmological origins for them have been proposed as well. Moreover, considering a power spectrum anisotropy due to the existence of a preferred spatial direction $\hat{n}$ and parametrized by a function $g(k)$ as 
\bea\label{disc}
P(\vec{k})=P(k)\left(1+g(k)(\hat{k}\cdot\hat{n})^2\right),
\eea
the five-year WMAP temperature data have been analyzed in order to find out what the magnitude and orientation of such an anisotropy could be. The magnitude has been found to be $g=0.29\pm 0.031$ and the orientation aligned nearly along the ecliptic poles \cite{Groeneboom:2009cb}. Similar results have been found in \cite{Hanson:2009gu}, where it is pointed out that the origin of such a signal is compatible with beam asymmetries (uncorrected in the maps) which should therefore be investigated before we can find out what the actual limits on the primordial $g$ are.  \\

\noindent Several fairly recent works have taken the direction of analysing the consequences, in terms of dynamics of the Universe and of cosmological fluctuations, of an anisotropic pre-inflationary or inflationary era. A cosmic no-hair conjecture exists according to which the presence of a cosmological constant at early times is expected to dilute any form of initial anisotropy \cite{Wald:1983ky}. This conjecture has been proven to be true for many (all Bianchi type cosmologies except for the the Bianchi type-IX, for which some restrictions are needed to ensure the applicability of the theorem), but not all other kinds of metrics and counterexamples exist in the literature \cite{Barrow:2005qv,Barrow:2006xb,DiGrezia:2003ug}. Moreover, even in the event isotropization should occur, there is a chance that signatures from anisotropic inflation or from an anisotropic pre-inflationary era might still be visible today \cite{Pereira:2007yy,Pitrou:2008gk,Dimastrogiovanni:2008ua,Gumrukcuoglu:2008gi}. In the same contest of searching for models of the early Universe that might produce some anisotropy signatures at late time, new theories have been proposed such as spinor models \cite{ArmendarizPicon:2003qk,Boehmer:2007ut,Watanabe:2009nc,deBerredoPeixoto:2009sb}, higher p-forms \cite{Germani:2009iq,Kobayashi:2009hj,Koivisto:2009ew,Germani:2009gg,Koivisto:2009fb,Koivisto:2009sd} and primordial vector field models (see Section~$2$ for a quick review). \\  

\noindent We are going to focus on statistical anisotropy and non-Gaussianity predictions of primordial vector field models. As mentioned above, there are great expectations that Planck and new experiments will, among other things, shed more light on the level of non-Gaussianity of the CMB fluctuations and on the nature of the unexpected anisotropy features we mentioned (see, e.g.,~\cite{Mandolesietal}). Models that combine both types of predictions could be more easily testable and, from non-Gaussianity measurement, more stringent statistical anisotropy predictions could be produced or viceversa.\\

\noindent Within vector field models, higher order correlators had been computed in \cite{Yokoyama:2008xw,Cogollo:2008bi,Dimopoulos:2008yv,Rodriguez:2008hy,Karciauskas:2008bc} and, more recently, in \cite{Dimopoulos:2009am,Dimopoulos:2009vu} for $U(1)$ vector fields. We considered $SU(2)$ vector field models in \cite{Bartolo:2009pa,Bartolo:2009kg}. Non-Abelian theories offer a richer amount of predictions compared to the Abelian case. Indeed, non-Abelian self interactions provide extra contributions to the bispectrum and trispectrum of curvature fluctuations that are naturally absent in the Abelian case. We verified that these extra contributions can be equally important in a large subset of the parameter space of the theory and, in some case, can even become the dominant ones. \\   

\noindent This paper is structured as follows: in Sec.~2 we review some vector field models of inflation; in Sec.~3 we present the $SU(2)$ model; in Sec.~4 we provide the results for the two, three and four-point functions of the curvature fluctuations; in Sec.~5 we present the non-Gaussianity amplitudes for the bispectrum and for the trispectrum; in Sec.~6 we show and discuss their shapes; finally in Sec.~7 we draw our conclusions.

\section{Inflation and primordial vector fields}

The attempt to explain some of the CMB ``anomalous'' features as the indication of a break of statistical isotropy is the main reason behind ours and many of the existing inflationary models populated by vector fields, but not the only one. The first one of these models \cite{Ford:1989me} was formulated with the goal of producing inflation by the action of vector fields, without having to invoke the existence of a scalar field. The same motivations inspired the works that followed \cite{Golovnev:2008cf,Golovnev:2008hv,Golovnev:2009ks}. Lately, models where primordial vector fields can leave an imprint on the CMB have been formulated as an alternative to the basic inflationary scenario, in the search for interesting non-Gaussianity predictions \cite{Yokoyama:2008xw,Cogollo:2008bi,Dimopoulos:2008yv,Rodriguez:2008hy,Karciauskas:2008bc,Bartolo:2009pa,Bartolo:2009kg,Dimopoulos:2009am,Dimopoulos:2009vu}. Finally, vector fields models of dark energy have been proposed \cite{ArmendarizPicon:2004pm,Boehmer:2007qa,Koivisto:2007bp,Koivisto:2008xf,Jimenez:2009py,Jimenez:2009zza}. All this appears to us as a rich bag of motivations for investigating these scenarios.\\
Before we quickly sketch some of them and list the results so far achieved in this direction, it is important to briefly indicate and explain the main issues and difficulties that these models have been facing. We will also shortly discuss the mechanisms of production of the curvature fluctuations in these models. \\
 
\noindent Building a model where primordial vector fields can drive inflation and/or produce the observed spectrum of large scale fluctuations requires a more complex Lagrangian than the basic gauge invariant $\mathcal{L}_{vector}=-(\sqrt{-g}/4) F^{\mu\nu}F_{\mu\nu}$. In fact, for a conformally invariant theory as the one described by $\mathcal{L}_{vector}$, vector fields fluctuations are not excited on superhorizon scales. It is then necessary to modify the Lagrangian. For some of the existing models, these modifications have been done to the expense of destabilizing the theory, by ``switching on'' unphysical degrees of freedom. This was pointed out in \cite{Himmetoglu:2008zp,Himmetoglu:2008hx,Himmetoglu:2009qi}, where a large variety of vector field models was analyzed in which longitudinal polarization modes exist that are endowed with negative squared masses (the ``wrong'' signs of the masses are imposed for the theory to satisfy the constraints that allow a suitable background evolution). It turnes out that, in a range of interest of the theory, these fields acquire negative total energy, i.e. behave like ``ghosts'', the presence of which is known to be responsible for an unstable vacuum. A related problem for some of these theories is represented by the existence of instabilities affecting the equations of motion of the ghost fields \cite{Himmetoglu:2008zp,Himmetoglu:2008hx,Himmetoglu:2009qi}.\\
In the remaing part of this section, we are going to present some of these models together with some recent attempts to overcome their limits. \\

\noindent In all of the models we will consider, primordial vector fields fluctuations end up either being entirely responsible for or only partially contributing to the curvature fluctuations at late times. This can happen through different mechanisms. If the vector fields affects the universe expansion during inflation, its contribution $\zeta_{A}$ to the total $\zeta$ can be derived from combining the definition of the number of e-foldings ($N=\int H dt$) with the Einstein equation ($H^2=(8 \pi G/3)(\rho_{\phi}+\rho_{A})$, $\rho_{A}$ being the energy density of the vector field and $\rho_{\phi}$ the inflaton energy density) and using the $\delta$N expansion of the curvature fluctuation in terms of both the inflaton and the vector fields fluctuations (see Sec.~4). To lowest order we have \cite{Dimopoulos:2008yv}
\bea\label{direct}
\zeta_{A}=\frac{A_{i}}{2 m_{P}^2}\delta A_{i},
\eea
where a single vector field has been taken into account for simplicity ($m_{P}$ is the reduced Planck mass, $A$ is the background value of the field and $\delta A$ its perturbation). When calculating the amplitude of non-Gaussianity in Sec.~5, we will refer to this case as ``vector inflation'' for simplicity.\\
A different fluctuation production process is the curvaton mechanism which was initially formulated for scalar theories but it is also applicable to vectors \cite{Dimopoulos:2006ms,Dimopoulos:2008rf}. Specifically, inflation is driven by a scalar field, whereas the curvaton field(s) (now played by the vectors), has a very small (compared to the Hubble rate) mass during inflation. Towards the end of the inflationary epoch, the Hubble rate value starts decreasing until it equates the vector mass; when this eventually happens, the curvaton begins to oscillate and it will then dissipate its energy into radiation. The curvaton becomes responsible for a fraction of the total curvature fluctuation that is proportional to a parameter ($r$) related to the ratio between the curvaton energy density and the total energy density of the universe at the epoch of the curvaton decay~\cite{Dimopoulos:2008yv} 
\bea\label{indirect}
\zeta_{A}=\frac{r}{3}\frac{\delta \rho_{A}}{\rho_{A}},
\eea
where $r\equiv 3\rho_{A}/(3\rho_{A}+4\rho_{\phi})$. Anisotropy bounds on the power spectrum favour small values of $r$.\\
From Eqs.~(\ref{direct}) and (\ref{indirect}) we can see that, dependending on which one of these two mechanisms of production of the curvature fluctuations is considered, different coefficients will result in the $\delta$N expansion (see Eq.~(\ref{expansion})).\\

\noindent In this section we will describe both models where inflation is intended to be vector field driven and those models in which, instead, the role of the inflaton is played by a scalar field, whereas the energy of the vector is kept subdominant in the total energy density of the universe during the entire inflationary phase.

\subsection{Self-coupled vector field models}

A pioneer work on vector field driven inflation was formulated by L.~H.~Ford \cite{Ford:1989me}, who considered a single self-coupled field $A_{\mu}$ with a Lagrangian
\bea
L_{vector}=-\frac{1}{4}F_{\mu\nu}F^{\mu\nu}+V(\psi)
\eea
where $F_{\mu\nu}\equiv \p_{\mu}B_{\nu}-\p_{\nu}B_{\mu}$ and the potential $V$ is a function of $\psi\equiv B_{\alpha}B^{\alpha}$. Different scenarios of expansion are analyzed by the author for different functions $V$. The universe expands anisotropically at the end of the inflationary era and this anisotropy either survives until late times or is damped out depending on the shape and the location of the minima of the potential.\\

\noindent The study of perturbations in a similar model was proposed by Dimopoulos in \cite{Dimopoulos:2006ms} where he showed that for a Lagrangian
\bea\label{dimopoulos}
L_{vector}=-\frac{1}{4}F_{\mu\nu}F^{\mu\nu}+\frac{1}{2}m^2 B_{\mu}B^{\mu}
\eea
and for $m^{2}\simeq -2H^2$, the transverse mode of the vector field is governed by the same equation of motion as a light scalar field in a de Sitter stage. A suitable superhorizon power spectrum of fluctuations could therefore arise. In order to prevent production of large scale anisotropy, in this model the vector field plays the role of the curvaton while inflation is driven by a scalar field. 

\subsection{Vector-field coupled to gravity}

The Lagrangian in Eq.~(\ref{dimopoulos}) may be also intended, at least during inflation, as including a non-minimal coupling of the vector field to gravity; indeed the mass term can be rewritten as
\bea\label{Lag}
L_{vector}\supset\frac{1}{2}\left(m^2_{0}+\xi R\right)B_{\mu}B^{\mu}
\eea
where, for the whole duration of the inflationary era, the bare mass $m_{0}$ is assumed to be much smaller than the Hubble rate and the Ricci scalar $R=-6\left[\frac{\ddot{a}}{a}+\left(\frac{\dot{a}}{a}\right)^2\right]$ can be approximated as $R\simeq-12 H^2$. For the specific value $\xi=1/6$, Eq.~(\ref{dimopoulos}) is retrieved.\\

\noindent For the Lagrangian just presented, Golovnev et al \cite{Golovnev:2008cf} proved that the problem of excessive anisotropy production in the case where inflation is driven by vector fields can be avoided if either a triplet of mutually orthogonal or a large number $N$ of randomly oriented vector fields is considered.\\

\noindent The Lagrangian (\ref{Lag}) with $\xi=1/6$ was also employed in \cite{Dimopoulos:2008rf}, where inflation is scalar-field-driven and a primordial vector field affects large-scale curvature fluctuations and, similarly, in \cite{Kanno:2008gn}, which includes a study of the backreaction of the vector field on the dynamics of expansion, by introducing a Bianchi type-I metric.

\subsection{Ackerman-Carroll-Wise (ACW) model}

A model was proposed in \cite{Ackerman:2007nb} where Lagrange multipliers ($\lambda$) are employed to determine a fixed norm primordial vector field $B_{\mu}B^{\mu}=m^2$
\bea\label{Acw}
L_{vector}\supset \lambda\left(B^{\mu}B_{\mu}-m^2\right)-\rho_{\Lambda}
\eea 
where $\rho_{\Lambda}$ is a vacuum energy. The expansion rate in this scenario is anisotropic: if we orient the $x$-axis of the spatial frame along the direction determined by the vector field, we find two different Hubble rates: along the $x$-direction it is equal to 
\bea
H_{b}^{2}=\frac{\rho_{\Lambda}}{m_{P}^{2}}\frac{1}{P(\mu)}, 
\eea
and it is given by $H_{a}=(1+c\mu^2)H_{b}$ along the orthogonal directions; $\mu\equiv m/m_{P}$, $P$ is a polynomial function of $\mu$ and $c$ is a parameter appearing in the kinetic part of the Lagrangian that we omitted in (\ref{Acw}) (see \cite{Ackerman:2007nb} for its complete expression). As expected, an isotropic expansion is recovered if the vev of the vector field is set to zero.

\subsection{Models with varying gauge coupling}

Most of the models mentioned so far successfully solve the problem of attaining a slow-roll regime for the vector-fields without imposing too many restrictions on the parameters of the theory and of avoiding excessive production of anisotropy at late times. None of them though escapes those instabilities related to the negative energy of the longitudinal modes (although a study of the instabilities for fixed-norm field models was done in \cite{Carroll:2009em} where some stable cases with non-canonical kinetic terms were found). As discussed in \cite{Himmetoglu:2008zp,Himmetoglu:2008hx,Himmetoglu:2009qi}, in the self-coupled model a ghost appears at small (compared to the horizon) wavelengths; in the non-minimally coupled and in the fixed-norm cases instead the instability concerns the region around horizon crossing.\\  

\noindent Models with varying gauge coupling can overcome the problem of instabilities and have recently attracted quite some attention. In \cite{Yokoyama:2008xw}, the authors consider a model of hybrid inflation \cite{Lyth:2005qk,Alabidi:2006wa,Salem:2005nd,Bernardeau:2004zz} with the introduction of a massless vector field
\bea
L\supset \frac{1}{2}\left(\p_{\mu}\phi\p^{\mu}\phi+\p_{\mu}\chi\p^{\mu}\chi\right)-\frac{1}{4}f^{2}(\phi)F_{\mu\nu}F_{\mu\nu}+V(\phi,\chi,B_{\mu})
\eea
where $\phi$ is the inflaton and $\chi$ is the so-called ``waterfall'' field. The potential $V$ is chosen in such a way as to preserve gauge invariance; this way the longitudinal mode disappears and instabilities are avoided.\\ 
Similarly, Kanno et al \cite{Watanabe:2009ct} consider a vector field Lagrangian of the type
\bea\label{}
L_{vector}=-\frac{1}{4}f^{2}(\phi)F^{\mu\nu}F_{\mu\nu},
\eea
but in a basic scalar field driven inflation model. Very recently, in \cite{DG} the linear perturbations in these kind of models have been investigated.\\
Finally, in \cite{Dimopoulos:2008yv,Dimopoulos:2009vu} varying mass vector field models have been introduced
\bea\label{var}
L_{vector}=-\frac{1}{4}f^{2}(\phi)F^{\mu\nu}F_{\mu\nu}+\frac{1}{2}m^{2}B_{\mu}B^{\mu},
\eea
where $f\simeq a^{\alpha}$ and $m\simeq a$ ($a$ is the scale factor and $\alpha$ is a numerical coefficient). The special cases $\alpha=1$ and $\alpha=-2$ are of special interest. In fact, introducing the fields $\tilde{A}_{\mu}$ and $A_{\mu}$, related to one another by $\tilde{A}_{\mu}\equiv f B_{\mu}=a A_{\mu}$ ($\tilde{A}_{\mu}$ and $A_{\mu}$ are respectively the comoving and the physical vectors), it is possible to verify that the physical gauge fields are governed by the same equations of motion as a light scalar field in a de Sitter background. Vector fields in this theory can then generate the observed (almost) scale invariant primordial power spectrum.

\section{$SU(2)$ vector model: equations of motion for the background and for linear perturbations}

Let us consider some models where inflation is driven by a scalar field in the presence of an $SU(2)$ vector multiplet \cite{Bartolo:2009pa,Bartolo:2009kg}. A fairly general Lagrangian can be the following
\bea\label{ac}\fl
S=\int d^{4}x \sqrt{-g}\left[\frac{m_{P}^{2}R}{2}-\frac{f^{2}(\phi)}{4}g^{\mu\al}g^{\nu\b}\sum_{a=1,2,3}F_{\mu\nu}^{a}F_{\al\b}^{a}-\frac{M^2}{2}g^{\mu\nu}\sum_{a=1,2,3}B_{\mu}^{a}B_{\nu}^{a}+L_{\phi}\right],
\eea
where $L_{\phi}$ is the Lagrangian of the scalar field and $F_{\mu\nu}^{a}\equiv\p_{\mu}B^{a}_{\nu}-\p_{\nu}B^{a}_{\mu}+g_{c}\ep^{abc}B^{b}_{\mu}B^{c}_{\nu}$ ($g_{c}$ is the $SU(2)$ gauge coupling). Both $f$ and the effective mass $M$ can be viewed as generic functions of time. The fields $B_{\mu}^{a}$ are comoving and related to the physical fields by $A_{\mu}^{a}=\left(B_{0}^{a},B_{i}^{a}/a\right)$. The free field operators can be Fourier expanded in their creation and annihilation operators
\bea\label{ip}
\fl
\delta A_{i}^{a}(\vec{x},\eta)=\int \frac{d^{3}q}{(2\pi)^{3}}e^{i\vec{q}\cdot\vec{x}}\sum_{\lambda=L,R,long}\Big[e^{\lambda}_{i}(\hat{q})a_{\vec{q}}^{a,\lambda}\delta A_{\lambda}^{a}(q,\eta)+e^{*\lambda}_{i}(-\hat{q})\left(a_{-\vec{q}}^{a,\lambda}\right)^{\dagger}\delta A_{\lambda}^{*a}(q,\eta)\Big],
\eea
where the polarization index $\lambda$ runs over left ($L$), right ($R$) and longitudinal ($long$) modes and
\bea
\left[a^{a,\lambda}_{\vec{k}},(a_{\vec{k}^{'}}^{a^{'},\lambda^{'}})^{\dagger}\right]=(2 \pi)^3\delta_{a,a^{'}}\delta_{\lambda,\lambda^{'}}\delta^{(3)}(\vec{k}-\vec{k}^{'}). 
\eea
Here $\eta$ the conformal time ($d\eta=dt/a(t)$). Once the functional forms of $f$ and $M$ have been specified, the equations of motion for the vector bosons can be written. For the most part, the calculations are quite general in this respect. In fact, the expression of all correlation functions, prior to explicitating the wavefunction for the gauge bosons, apply to any $SU(2)$ theory with an action as in (\ref{ac}), both for what we will call the ``Abelian'' and for the ``non-Abelian'' contributions. In particular, the structure of the interaction Hamiltonian is independent of the functional dependence of $f$ and $M$ and determines the general form of and the anisotropy coefficients appearing in the final ``non-Abelian'' expressions (see Sec.~4). When it comes to explicitate the wavefunctions, a choice that can help keeping the result as easy to generalize as possible is the following 
\bea\label{T}
\delta B^{T}=-\frac{\sqrt{\pi x}}{2\sqrt{k}}\left[J_{3/2}(x)+iJ_{-3/2}(x)\right],
\eea
for the transverse mode and
\bea\label{L}
\delta B^{||}=n(x)\delta B^{T}\, ,
\eea
for the longitudinal mode ($n$ is a unknown function of $x\equiv -k \eta$) \cite{Bartolo:2009pa,Bartolo:2009kg}. Let us see why. As previously stated, for $f \simeq a^{\alpha}$ and with $\alpha=0,1,-2$, it is possible to verify that the (physical) transverse mode behaves exactly like a light scalar field in a de Sitter background. Considering the solution (\ref{T}) then takes into account at least these special cases. As to the longitudinal mode, a parametrization was adopted as in (\ref{L}) in order to keep the analysis more general and given that, because of the instability issues, introducing this degree of freedom into the theory requires special attention. We are going to keep the longitudinal mode ``alive'' in the calculations we present, by considering a nonzero function $n(x)$, and focus on the simplest case of $f=1$. This case is known to be affected by quantum instabilities in the longitudinal mode, anyway we choose $f=1$ for the sake of simplicity in our presentation. The results can be easily generalized to gauge invariant models (please refer to \cite{Bartolo:2009kg} for a sample generalization of the calculations to massless $f\simeq a^{(1,-2)}$ models).

\section{Correlation functions of curvature fluctuations: analytic expressions}

We are now ready to review the computation of the power spectrum, bispectrum and trispectrum for the curvature fluctuations $\zeta$ generated during inflation
\bea\label{ps}
\langle\zeta_{\vec{k}_{{1}}}\zeta_{\vec{k}_{{2}}}\rangle=(2 \pi)^3\delta^{(3)}(\vec{k_{1}}+\vec{k_{2}})P_{\zeta}(\vec{k}),
\\
\label{bisp}
\langle\zeta_{\vec{k}_{{1}}}\zeta_{\vec{k}_{{2}}}\zeta_{\vec{k}_{{3}}} \rangle=(2 \pi)^3\delta^{(3)}(\vec{k_{1}}+\vec{k_{2}}+\vec{k_{3}})B_{\zeta}(\vec{k_{1}},\vec{k_{2}},\vec{k_{3}})\\
\label{trisp}
\langle\zeta_{\vec{k}_{1}}\zeta_{\vec{k}_{2}}\zeta_{\vec{k}_{3}}\zeta_{\vec{k}_{4}} \rangle =(2 \pi)^3\delta^{(3)}(\vec{k}_{1}+\vec{k}_{2}+\vec{k}_{3}+\vec{k}_{4})T_{\zeta}(\vec{k}_{1},\vec{k}_{2},\vec{k}_{3},\vec{k}_{4}).
\eea
Notice that, on the right-hand side of (\ref{ps}) through (\ref{trisp}), we indicated a dependence from the direction of the wavevectors; in models of inflation where isotropy is preserved, the power spectrum and the bispectrum only depend on the moduli of the wave vectors. This will not be the case for the $SU(2)$ model.\\
\noindent The $\delta$N formula \cite{deltaN1,deltaN2,deltaN3,deltaN4} will be employed
\bea\label{deltaN}
\zeta(\vec{x},t)=N(\vec{x},t^{*},t)-N(t^{*},t)\equiv\delta N(\vec{x},t),
\eea
which holds if times $t^{*}$ and $t$ are chosen, respectively, on a flat and on a uniform density temporal slices ($N$ is the number of e-foldings of inflation occurring between these two times)~\footnote{We employ a spatial metric $g_{ij}=a^{2}(t)e^{-2\Psi}\left(e^{\gamma}\right)_{ij}$, and at linear level the curvature perturbation corresponds to
$\zeta\equiv -\Psi+H \delta u$, where $\delta u$ is the fluctuation in the total energy density and $\Psi$ is a scalar metric fluctuation.}. 
In the presence of a single scalar field $\phi$, Eq.~(\ref{deltaN}) is further expandable as
\bea\label{deltaN2}
\zeta(\vec{x},t)=\sum_{n}\frac{N^{(n)}(t^{*},t)}{n!}\left(\delta \phi(\vec{x},t^{*})\right)^{n}
\eea
where $N^{(n)}$ is the  partial derivative of the e-folding number w.r.t. $\phi$ on the initial hypersurface $t^{*}$.\\ If we apply (\ref{deltaN2}) to the inflaton$+SU(2)$vector model, we have
\bea
\label{expansion}\fl
\zeta(\vec{x},t)&=& N_{\phi}\df+N^{\mu}_{a}\delta A_{\mu}^{a}+\frac{1}{2}N_{\phi\phi}\left(\df\right)^2+\frac{1}{2}N^{\mu\nu}_{ab}\delta A_{\mu}^{a}\delta A_{\nu}^{b}+N_{\phi a}^{\mu}\df\delta A_{\mu}^{a}\nonumber\\\fl&+&\frac{1}{3!}N_{\phi\phi\phi}(\df)^3+\frac{1}{3!}N_{abc}^{\mu\nu\lambda}\de A_{\mu}^{a} \de A_{\nu}^{b} \de A_{\lambda}^{c}+\frac{1}{2}N_{\phi\phi a}^{\mu}(\df)^2\de A_{\mu}^{a}+\frac{1}{2}N_{\phi ab}^{\mu\nu}\df\de A_{\mu}^{a}\de A_{\nu}^{b}\nonumber\\\fl&+&\frac{1}{3!}N_{\phi\phi\phi\phi}(\df)^4+\frac{1}{3!}N_{abcd}^{\mu\nu\lambda\eta}\de A_{\mu}^{a}\de A_{\nu}^{b}\de A_{\lambda}^{c}\de A_{\eta}^{d}+...,
\eea
where now
\bea\label{NNN}
N_{\phi}\equiv \left(\frac{\p N}{\p \phi}\right)_{t^{*}},\,\,\,\,\,
N^{\mu}_{a}\equiv\left(\frac{\p N}{\p A^{a}_{\mu}}\right)_{t^{*}},\,\,\,\,\,N_{\phi a}^{\mu}\equiv\left(\frac{\p^{2} N}{\p \phi\p A^{a}_{\mu}}\right)_{t^{*}}
\eea
and so on for higher order derivatives.\\ Our plan is to show the derivation the correlation functions of $\zeta$ from the ones of $\df$ and $\delta A_{i}^{a}$, after a replacement of the $\delta$N expansion (\ref{expansion}) in Eqs.~(\ref{ps}) through (\ref{trisp}). \\
The correlation functions can be evaluated using the Schwinger-Keldysh formula \cite{Schwinger:1960qe,Calzetta:1986ey,Jordan:1986ug,Weinberg:2005vy,Weinberg:2006ac}
\bea\label{sk}
\langle\Omega|\Theta(t)|\Omega\rangle=\left\langle 0\left|\left[\bar{T}\left(e^{i {\int}^{t}_{0}H_{I}(t')dt'}\right)\right]\Theta_{I}(t)\left[T \left(e^{-i {\int}^{t}_{0}H_{I}(t')dt'}\right)\right]\right|0\right\rangle,
\eea
where, on the left-hand side, the operator $\Theta$ and the vacuum $\Omega$ are in the interacting theory whereas, on the right-hand side, all operators are in the so-called ``interaction picture'', i.e. they can be treated as free fields (the Fouries expansion in Eq.~(\ref{ip}) thus apply), and $|0\rangle$ is the free theory vacuum.\\

\noindent When calculating the spectra of $\zeta$, the perturbative expansions in Eq.~(\ref{expansion}) and (\ref{sk}) will be carried out to only include tree-level contributions, neglecting higher order ``loop'' terms, either classical, i.e. from the $\delta$N series, or of quantum origin, i.e. from the Schwinger-Keldysh series. Assuming that the $SU(2)$ coupling $g_{c}$ is ``small'' and that we are dealing with ``small'' fluctuations in the fields and given the fact that a slow-roll regime is being assumed, it turns out that it is indeed safe for the two expansions to be truncated at tree-level.\\ 

\noindent The correlation functions of $\zeta$ will then result as the sum of scalar, vector and (scalar and vector) mixed contributions. As to the vector part, this will be made up of terms that are merely generated by the $\delta$N expansion, i.e. they only include the zeroth order of the in-in formula (we call these terms ``Abelian'', being them retrievable in the $U(1)$ case), and by (``non-Abelian'') terms arising from the Schwinger-Keldysh operator expansion beyond zeroth order, i.e. from the gauge fields self-interactions.\\

\noindent Let us now discuss the level of generality of the results we will present in the next sections, w.r.t. the choice of a specific Lagrangian.\\ The expression for the Abelian contributions provided in Secs,~$4.1$ and $4.2.1$, apply to any $SU(2)$ model of gauge interactions with no direct coupling between scalar and vector fields (extra terms would be otherwise needed in Eqs.~(\ref{class1}) and (\ref{class2})). The next stage in the Abelian contributions computation would be to explicitate the derivatives of the e-foldings number and the wavefunctions of the fields: they both depend on the equations of motion of the system, therefore the fixing of a specific model is required at this point.\\
As to the non-Abelian contributions, the results in Eqs.~(\ref{class3}) and (\ref{class4}) are completely general except for assuming, again, that no direct vector-scalar field coupling exists. The structure of Eqs.~(\ref{fire}) and (\ref{fire1}) is instead due to the choice of a non-Abelian gauge group. The expressions of the anisotropy coefficients $I_{n}$ and $L_{n}$ in Eqs.~(\ref{fire}) and (\ref{fire1}) depend on the specific non-Abelian gauge group (for $SU(2)$ one of the $I_{n}$ is given in Eq.~(\ref{sample2})). Finally, the specific expressions of the isotropic functions $F_{n}$ (a sample of which is shown in Eq.~(\ref{sample1})) and $G_{n}$ were derived considering the Lagrangian (\ref{ac}) with $f=1$ and the eigenfunctions for the vector bosons provided in Eqs.~(\ref{T}) and (\ref{L}).

\subsection{The power spectrum}

The power spectrum of $\zeta$ can be straightforwardly derived at tree-level, using the $\delta$N expansion (\ref{expansion}), from the inflaton and the vector fields power spectra
\bea\label{power-zeta}
P_{\zeta}(\vec{k})&=&P^{iso}(k)\left[1+g^{ab}\left(\hat{k}\cdot\hat{N}_{a}\right)\left(\hat{k}\cdot\hat{N}_{b}\right)+is^{ab}\hat{k}\cdot\left(\hat{N}_{a}\times\hat{N}_{b}\right)\right].
\eea
The isotropic part of the previous expression has been factorized in 
\bea\label{piso}
P^{iso}(k)\equiv N^{2}_{\phi}P_{\phi}(k)+\left(\vec{N}_{c}\cdot\vec{N}_{d}\right)P^{cd}_{+},
\eea
where we have defined the following combinations
\bea
P_{\pm}^{ab}\equiv (1/2)(P_{R}^{ab}\pm P_{L}^{ab}),
\eea 
from the power spectra for the right, left and longitudinal polarization modes
\bea
P_{R}^{ab}&\equiv& \delta_{ab}\delta A_{R}^{a}(k,t^{*})\delta A_{R}^{b*}(k,t^{*}),\\
P_{L}^{ab}&\equiv& \delta_{ab}\delta A_{L}^{a}(k,t^{*})\delta A_{L}^{b*}(k,t^{*}),\\
P_{long}^{ab}&\equiv& \delta_{ab}\delta A_{long}^{a}(k,t^{*})\delta A_{long}^{b*}(k,t^{*})
\eea
The anisotropic parts are weighted by the coefficients
\bea
g^{ab}\equiv \frac{N^{a}N^{b}\left(P^{ab}_{long}-P^{ab}_{+}\right)}{N^{2}_{\phi}P_{\phi}+\left(\vec{N}_{c}\cdot\vec{N}_{d}\right)P^{cd}_{+}},\\
s^{ab}\equiv \frac{N^{a}N^{b}P^{ab}_{-}}{N^{2}_{\phi}P_{\phi}+\left(\vec{N}_{c}\cdot\vec{N}_{d}\right)P^{cd}_{+}},
\eea
(where a sum is intended over indices $c$ and $d$ but not over $a$ and $b$). Eq.~(\ref{piso}) can also be written as
\bea
P^{iso}(k)=N^{2}_{\phi}P_{\phi}\left[1+\beta_{cd}\frac{P^{cd}_{+}}{P_{\phi}}\right],
\eea
after introducing the parameter
\bea\label{beta}
\beta_{cd}\equiv \frac{\vec{N}_{c}\cdot\vec{N}_{d}}{N^{2}_{\phi}}.
\eea
Notice that what when we say ``isotropic'', as far as the expression for the power spectrum is concerned, we simply mean ``independent'' of the direction of the wave vector. In this case instead, the vector bosons introduce three preferred spatial directions: the r.h.s. of Eq.~(\ref{power-zeta}) depends on their orientation w.r.t. the wave vector.\\

\noindent As expected, the coefficients $g^{ab}$ and $s^{ab}$ that weight the anisotropic part of the power spectrum are related to $\beta_{cd}$, i.e. to the parameters that quantify how much the expansion of the universe is affected by the vector bosons compared to the scalar field.\\
Assuming no parity violation in the model, we have $s^{ab}=0$; the parameters $g^{ab}$ and $\beta_{ab}$ are instead unconstrained. In the $U(1)$ case and for parity conserving theories, Eq.~(\ref{power-zeta}) reduces to \cite{Dimopoulos:2008yv}
\bea\label{pdisc}
P_{\zeta}(\vec{k})=P^{iso}_{\zeta}(k)\left[1+g\left(\hat{k}\cdot\hat{n}\right)\right]
\eea
where $\hat{n}$ indicates the preferred spatial direction; also one can check that in this simple case, if $P_{+}\simeq P_{\phi}$ and $P_{long}=k P_{+}$ ($k\neq 1$), the relation $g=(k-1)\beta/(1+\beta)$ holds, where $\beta\equiv \left(N_{A}/N_{\phi}\right)^2$ (the anisotropy coefficient $g$ is not to be confused with the $SU(2)$ coupling constant $g_{c}$). If it is safe to assume $|g| \ll 1$ (see discussion following Eq.~(\ref{disc}) and references \cite{Groeneboom:2009cb,Hanson:2009gu}), a similar upper bound can also be placed on $\beta$.  \\
In the case where more than one special directions exists, as in the $SU(2)$ model, no such analysis on the anisotropy data has been so far carried out, the $g^{a}$ parameters cannot then be constrained, unless assuming that the three directions converge into a single one; in that case a constraint could be placed on the sum $|g|\equiv |\sum_{a}g^{a}|$, where $a=1,2,3$ and $P_{\zeta}(\vec{k})=P^{iso}_{\zeta}(k)\left[1+g^{a}\left(\hat{k}\cdot\hat{n}_{a}\right)\right]$.

\subsection{Higher-order correlators}

We will present the results for the tree-level contributions to the bispectrum and to the trispectrum of $\zeta$. \\
These can be classified in two cathegories, that we indicate as ``Abelian'' and ``non-Abelian''. The former are intended as terms that merely arise from the $\delta$N expansion and are thus retrievable in the Abelian case; the latter are derived from the linear and quadratic expansions (in terms of the gauge bosons interaction Hamiltonian) of the Schwinger-Keldysh formula and are therefore peculiar to the non-Abelian case.\\
We are going to provide both types of contributions, in preparation for discussing and comparing their magnitudes later on in Sec.~5.

\subsubsection{Abelian contributions}
${}$\\
By plugging the $\delta$N expansion (\ref{expansion}) in Eqs.~(\ref{bisp}) and (\ref{trisp}), we have
\bea\label{class1}\fl
B_{\zeta}(\vec{k_{1}},\vec{k_{2}},\vec{k_{3}})&\supset&\frac{1}{2}N_{\phi}^{2}N_{\phi\phi}\left[P_{\phi}({k}_{1})P_{\phi}({k}_{2})+perms.\right]\nonumber\\\fl&+&\frac{1}{2}N_{a}^{\mu}N_{b}^{\nu}N_{cd}^{\rho\sigma}\left[\Pi^{ac}_{\mu\rho}(\vec{k}_{1})\Pi^{bd}_{\nu\sigma}(\vec{k}_{2})+perms.\right]
\nonumber\\\fl&+&\frac{1}{2}N_{\phi}N_{a}^{\mu}N_{\phi b}^{\nu}\left[P_{\phi}({k}_{1})\Pi^{ab}_{\mu\nu}(\vec{k}_{2})+perms.\right]\nonumber\\\fl&+&N_{\phi}N_{\phi}^{2}B_{\phi}(k_{1},k_{2},k_{3}),
\eea
for the bispectrum and
\bea\label{class2}\fl
T_{\zeta}(\vec{k_{1}},\vec{k_{2}},\vec{k_{3}},\vec{k_{4}})&\supset&N_{\phi}^{4}T_{\phi}(\vec{k}_{1},\vec{k}_{2},\vec{k}_{3},\vec{k}_{4})\nonumber\\\fl&+&N_{\phi}^{3}N_{\phi\phi}\left[P_{\phi}(k_{1})B_{\phi}(|\vec{k}_{1}+\vec{k}_{2}|,k_{3},k_{4})+perms.\right]\nonumber\\&+&N_{\phi}^{2}N_{a}^{\mu}N_{\phi b}^{\nu}\left[P_{\mu\nu}^{ab}(\vec{k}_{3})B_{\phi}^{}(k_{1},k_{2},|\vec{k}_{3}+\vec{k}_{4}|)+perms.\right]\nonumber\\\fl&+&N_{\phi}^{2}N_{\phi\phi}^{2}\left[P_{\phi}(k_{1})P_{\phi}(k_{2})P_{\phi}(|\vec{k}_{1}+\vec{k}_{3}|)+perms.\right]\nonumber\\&+&N_{\phi}^{3}N_{\phi\phi\phi}\left[P_{\phi}({k}_{1})P_{\phi}(k_{2})P_{\phi}(k_{3})+perms.\right]\nonumber\\\fl
&+&N_{\phi}^{2}N_{\phi a}^{\mu}N_{\phi b}^{\nu}\left[P_{\mu\nu}^{ab}(\vec{k}_{1}+\vec{k}_{3})P_{\phi}^{}({k}_{1})P_{\phi}^{}({k}_{2})+perms.\right]\nonumber\\\fl&+&N_{a}^{\mu}N_{b}^{\nu}N_{\phi c}^{\rho}N_{\phi d}^{\sigma}\left[P_{\mu\rho}^{ac}(\vec{k}_{1})P_{\nu\sigma}^{bd}(\vec{k}_{2})P_{\phi}^{}(|\vec{k}_{1}+\vec{k}_{3}|)+perms.\right]\nonumber\\\fl&+&N_{\phi}^{2}N_{a}^{\mu}N_{\phi\phi b}^{\nu}\left[P_{\phi}^{}({k}_{1})P_{\phi}^{}({k}_{2})P_{\mu\nu}^{ab}(\vec{k}_{3})+perms.\right]\nonumber\\\fl&+&N_{\phi}^{}N_{a}^{\mu}N_{b}^{\nu}N_{\phi cd}^{\rho\sigma}\left[P_{\mu\rho}^{ac}(\vec{k}_{1})P_{\nu\sigma}^{bd}(\vec{k}_{2})P_{\phi}^{}({k}_{3})+perms.\right]\nonumber\\\fl&+&N_{\phi\phi}N_{\phi}N_{\phi a}^{\mu}N_{b}^{\nu}\left[P_{\phi}(k_{2})P_{\phi}(|\vec{k}_{1}+\vec{k}_{2}|)P_{\mu\nu}^{ab}(\vec{k}_{4})+perms.\right]\nonumber\\\fl&+&N_{ab}^{\mu\nu}N_{c}^{\rho}N_{\phi d}^{\sigma}N_{\phi}\left[P_{ac}^{\mu\rho}(\vec{k}_{2})P_{bd}^{\nu\sigma}(\vec{k}_{1}+\vec{k}_{2})P_{\phi}(k_{4})+perms.\right]\nonumber\\\fl&+&N_{a}^{\mu}N_{b}^{\nu}N_{cd}^{\rho\sigma}N_{ef}^{\delta\eta}\left[P_{\mu\rho}^{ac}(\vec{k}_{1})P_{\nu\delta}^{be}(\vec{k}_{2})P_{\sigma\eta}^{df}(\vec{k}_{1}+\vec{k}_{3})+perms.\right]\nonumber\\&+&N_{a}^{\mu}N_{b}^{\nu}N_{c}^{\rho}N_{def}^{\sigma\delta\eta}\left[P_{\mu\sigma}^{ad}(\vec{k}_{1})P_{\nu\delta}^{be}(\vec{k}_{2})P_{\rho\eta}^{cf}(\vec{k}_{3})+perms.\right], 
\eea
for the trispectrum.\\ Before we proceed with explicitating these quantities and for the rest of the paper, the $N^{a}_{0}$ coefficients will be set to zero. This choice was discussed in Sec.~2 and, more in details, in Appendix A of \cite{Bartolo:2009pa}. Summarizing, it is possible to verify that the temporal mode $B_{0}^{a}=0$ is a solution to the equations of motion for the vector bosons, after slightly restricting the parameter space of the theory. The adoption of this kind of solution, which is related to the assumption of a slow-roll regime for the vector fields, implies that the derivatives of $N$ w.r.t. the temporal mode can be set to zero.\\
\noindent Let us now provide some definition for the quantities introduced in (\ref{class1})-(\ref{class2}): we are going to switch from the greek indices $\mu$, $\nu$, ... to the latin ones, generally used for labelling the three spatial directions, in order to stress that all of the vector quantities will be from now on three-dimensional
\bea
\Pi_{ij}^{ab}(\vec{k}) \equiv T^{even}_{ij}(\vec{k})P_{+}^{ab}+i T^{odd}_{ij}(\vec{k})P_{-}^{ab}+T^{long}_{ij}(\vec{k})P_{long}^{ab},
\eea 
where
\bea
T^{even}_{ij}(\vec{k}) &\equiv& e_{i}^{L}(\hat{k})e_{j}^{*L}(\hat{k})+e_{i}^{R}(\hat{k})e_{j}^{*R}(\hat{k}),\\
T^{odd}_{ij}(\vec{k}) &\equiv& i \Big[e_{i}^{L}(\hat{k})e_{j}^{*L}(\hat{k})-e_{i}^{R}(\hat{k})e_{j}^{*R}(\hat{k})\Big],\\
T^{long}_{ij}(\vec{k}) &\equiv& e_{i}^{l}(\hat{k})e_{j}^{*l}(\hat{k}).
\eea
The polarization vectors are $e^{L}(\hat{k})\equiv\frac{1}{\sqrt{2}} (\cos\theta\cos\phi-i\sin\phi,\cos\theta\sin\phi+i\cos\phi,-\sin\theta)$, $e^{R}(\hat{k})=e^{*L}(\hat{k})$ and $e^{l}(\hat{k})=\hat{k}=(\sin\theta\cos\phi,\sin\theta\sin\phi,\cos\theta)$, from which we have
\bea
T^{even}_{ij}(\vec{k})&=&\delta_{ij}-\hat{k}_{i}\hat{k}_{j},\\
T^{odd}_{ij}(\vec{k})&=&\epsilon_{ijk}\hat{k}_{k},\\
T^{long}_{ij}(\vec{k})&=&\hat{k}_{i}\hat{k}_{j}.
\eea
The purely scalar terms in Eqs.~(\ref{class1})-(\ref{class2}) are already known from the literature \footnote{In single-field slow-roll inflation $P_{\phi}=H_{*}^{2}/2k^3$, where $H_{*}$ is the Hubble rate evaluated at horizon exit; the bispectrum and the trispectrum of the scalar field ($B_{\phi}$ and $T_{\phi}$) can be found in \cite{Acqua,Maldacena:2002vr,Seery:2005wm,Seery:2006vu,Seery:2008ax} (they were also reported in Eqs.~(11) and (12) of \cite{Bartolo:2009kg}).}. As to the mixed (scalar-vector) terms, they can be ignored if one considers a Lagrangian where there is no direct coupling between the inflaton and the gauge bosons and where slow-roll assumptions are introduced for the fields (see Sec.~4 of \cite{Bartolo:2009kg} for a complete discussion on this). Let us then look at the (purely) vector part. Its anisotropy features can be stressed by rewriting them as follows 
\bea\label{b}\fl
B_{\zeta}(\vec{k_{1}},\vec{k_{2}},\vec{k_{3}})&\supset&\frac{1}{2}N_{a}^{i}N_{b}^{j}N_{cd}^{kl}\Pi^{ac}_{ik}(\vec{k}_{1})\Pi^{bd}_{jl}(\vec{k}_{2})=M^{c}_{k} N^{kl}_{cd} M^{d}_{l}\\\label{t}\fl
T_{\zeta}(\vec{k_{1}},\vec{k_{2}},\vec{k_{3}},\vec{k_{4}})&\supset&N_{a}^{\mu}N_{b}^{\nu}N_{cd}^{\rho\sigma}N_{ef}^{\delta\eta}P_{\mu\rho}^{ac}(\vec{k}_{1})P_{\nu\delta}^{be}(\vec{k}_{2})P_{\sigma\eta}^{df}(\vec{k}_{1}+\vec{k}_{3})\nonumber\\\fl&+&N_{a}^{\mu}N_{b}^{\nu}N_{c}^{\rho}N_{def}^{\sigma\delta\eta}P_{\mu\sigma}^{ad}(\vec{k}_{1})P_{\nu\delta}^{be}(\vec{k}_{2})P_{\rho\eta}^{cf}(\vec{k}_{3})\nonumber\\\fl&=&M_{i}^{c}L_{ce}^{ij}M_{j}^{e}+M_{i}^{f}M_{j}^{e}M_{k}^{d}N_{fed}^{ijk},
\eea
where
\bea\label{nonzero2}\fl
M_{k}^{c}(\vec{k})&\equiv& N_{a}^{i}P_{ik}^{ac}(\vec{k})=P^{ac}_{+}(k)\left[\delta_{ik}{N}_{a}^{i}+p^{ac}(k)\hat{k}_{k}\left(\hat{k}\cdot\vec{N}_{a}\right)+iq^{ac}(k)\left(\hat{k}\times\vec{N}_{a}\right)_{k}\right]\\\fl
L_{ce}^{jl}(\vec{k})&\equiv& N_{cd}^{ji}P_{ik}^{df}(\vec{k})N_{fe}^{kl}\nonumber\\\fl&=&P_{+}^{df}(\vec{k})\big[\vec{N}_{cd}^{j}\cdot\vec{N}_{ef}^{l}+p^{df}(k)\left(\hat{k}\cdot\vec{N}_{cd}^{j}\right)\left(\hat{k}\cdot\vec{N}_{ef}^{l}\right)+iq^{df}(k)\hat{k}\cdot\vec{N}^{j}_{cd}\times \vec{N}_{ef}^{l}\big]. 
\eea
In the previous equations, we defined
\bea
p^{ac}(k)&\equiv& \frac{P^{ac}_{long}-P^{ac}_{+}}{P^{ac}_{+}},\\
q^{ac}(k) &\equiv& \frac{P^{ac}_{-}}{P^{ac}_{+}},
\eea
with $\vec{N}_{a}\equiv (N_{a}^{1},N_{a}^{2},N_{a}^{3})$ and $\vec{N}_{cd}^{j}\equiv (N_{cd}^{j1},N_{cd}^{j2},N_{cd}^{j3})$.\\
Notice that, as for the power spectrum (\ref{power-zeta}), also in Eqs.~(\ref{b})-(\ref{t}) the anisotropic parts of the expressions are weighted by coefficients that are proportional either to $P_{-}$ or to $(P_{long}-P_{+})$. When these two quantities are equal to zero, the (Abelian) bispectrum and trispectrum are therefore isotropized. $P_{-}=0$ in parity conserving theories, like the ones we have been describing. According to the parametrization (\ref{L}) of the longitudinal mode, we have $P_{long}-P_{+}=(|n(x)|^2-1) P_{+}$.

\subsubsection{Non-Abelian contributions}
${}$\\
We list the non-Abelian terms for the bispectrum 
\bea\label{class3}\fl
B_{\zeta}(\vec{k_{1}},\vec{k_{2}},\vec{k_{3}})&\supset& N_{a}^{i}N_{b}^{j}N_{c}^{k}B_{ijk}^{abc}(\vec{k}_{1},\vec{k}_{2},\vec{k}_{3})
\eea
and for the trispectrum
\bea\label{class4}\fl
T_{\zeta}(\vec{k_{1}},\vec{k_{2}},\vec{k_{3}},\vec{k_{4}})&\supset& N_{a}^{i}N_{b}^{j}N_{c}^{k}N_{d}^{l}T_{ijkl}^{abcd}(\vec{k}_{1},\vec{k}_{2},\vec{k}_{3},\vec{k}_{4})\nonumber\\&+&N_{a}^{i}N_{b}^{j}N_{\phi}N_{\phi c}^{k}\left[P_{\phi}^{}({k}_{3})B_{ijk}^{abc}(\vec{k}_{1},\vec{k}_{2},\vec{k}_{3}+\vec{k}_{4})+perms.\right]\nonumber\\&+&N_{a}^{i}N_{b}^{j}N_{c}^{k}N_{de}^{lm}\left[P_{il}^{ad}(\vec{k}_{1})B_{jkm}^{bce}(\vec{k}_{1}+\vec{k}_{2},\vec{k}_{3},\vec{k}_{4})+perms.\right].
\eea
The computation of the vector bosons spectra
\bea
\langle \de A^{a}_{i} \de A^{b}_{j} \de A^{c}_{k}\rangle=\delta^{(3)}(\vec{k}_{1}+\vec{k}_{2}+\vec{k}_{3})B_{ijk}^{abc},\\
\langle \de A^{a}_{i} \de A^{b}_{j} \de A^{c}_{k} \de A^{d}_{l}\rangle=\delta^{(3)}(\vec{k}_{1}+\vec{k}_{2}+\vec{k}_{3}+\vec{k}_{4})T^{abcd}_{ijkl}, 
\eea
will be reviewed in this section. This requires the expansion of the in-in formula up to second order in the interaction Hamiltonian
\bea\fl
\langle\Theta(\eta^{*})\rangle &\supset& i\langle T \Big[\Theta \int_{-\infty}^{\eta^{*}}d\eta^{'}\left(H^{+}_{int}(\eta^{'})-H^{-}_{int}(\eta^{'})\right)\Big]\rangle\\\fl&+&
\frac{(-i)^2}{2}\langle T \Big[\Theta \int_{-\infty}^{\eta^{*}}d\eta^{'}\left(H^{+}_{int}(\eta^{'})-H^{-}_{int}(\eta^{'})\right) \int_{-\infty}^{\eta^{*}}d\eta^{''}\left(H^{+}_{int}(\eta^{''})-H^{-}_{int}(\eta^{''})\right)\Big]\rangle.\nonumber
\eea

\noindent The interaction Hamiltonian needs to be expanded up to fourth order in the fields fluctuations, i.e. $H_{int}=H_{int}^{(3)}+H_{int}^{(4)}$, where
\bea
\label{int3}
H_{int}^{(3)}&=&g_{c}\ep^{abc}g^{ik}g^{jl}\left(\p_{i}\DB_{j}^{a}\right)\DB^{b}_{k}\DB_{l}^{c}+g_{c}^2 \ep^{eab}\ep^{ecd}g^{ik}g^{jl}B^{a}_{i}\de B^{b}_{j}\de B^{c}_{k}\de B^{d}_{l}\\\label{int4}
H_{int}^{(4)}&=&g_{c}^2 \ep^{eab}\ep^{ecd}g^{ij}g^{kl}\de B^{a}_{i}\de B^{b}_{k}\de B^{c}_{j}\de B^{d}_{l}.
\eea

\begin{figure}\centering
 \includegraphics[width=0.3\textwidth]{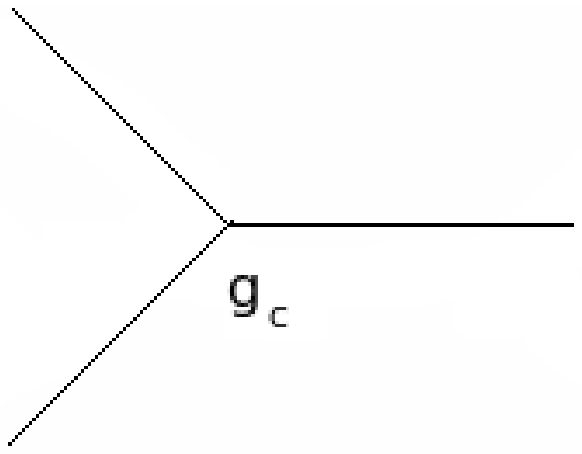}
\hspace{0.17\textwidth}
 \includegraphics[width=0.3\textwidth]{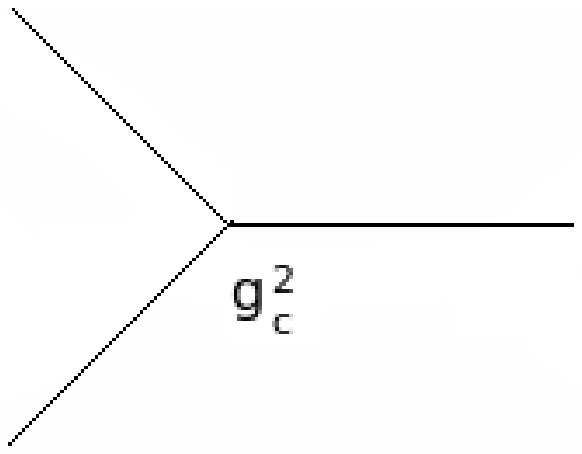}
\caption{ \label{Fig1} Diagrammatic representations of the tree-level contributions to the vector fields bispectrum.}
\end{figure}

\begin{figure}\centering
 \includegraphics[width=0.3\textwidth]{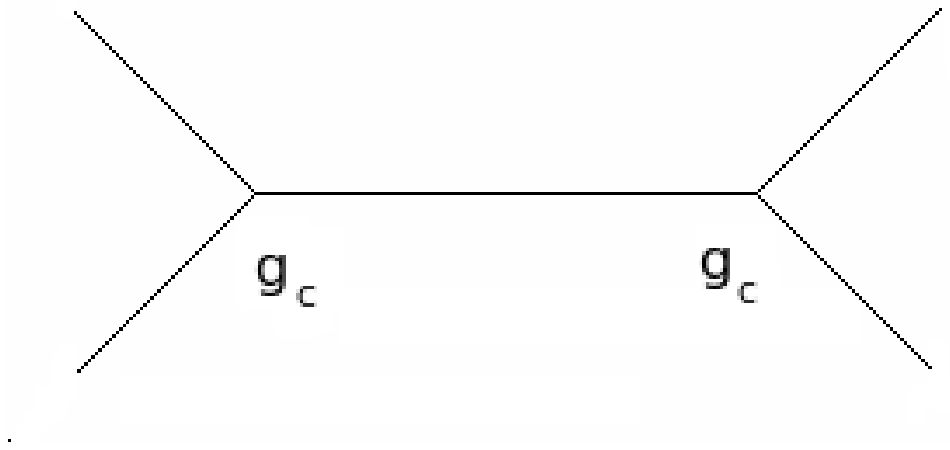}
\hspace{0.17\textwidth}
 \includegraphics[width=0.3\textwidth]{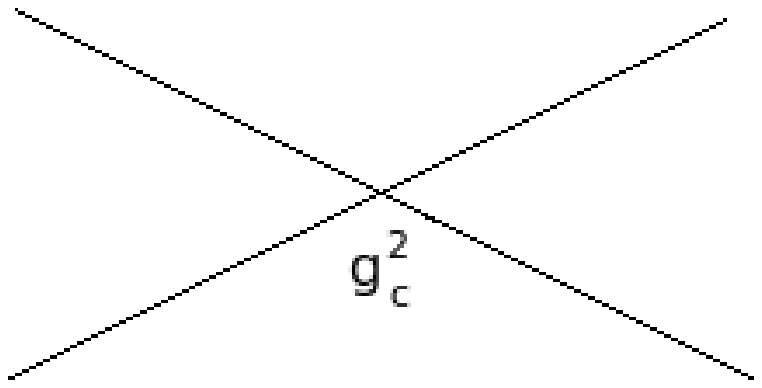}
\caption{ \label{Fig1} Diagrammatic representations of the tree-level contributions to the vector fields trispectrum: vector-exchange (on the left) and contact-interaction (on the right) diagrams.}
\end{figure}

\noindent To tree-level, the relevant diagrams are pictured in Figs.~1 and 2. By looking at Eqs.~(\ref{int3}) and (\ref{int4}), we can see that there is a bispectrum diagram that is lower in terms of power of the $SU(2)$ coupling ($\sim g_{c}$) compared to the trispectrum ($\sim g_{c}^2$); as a matter of fact, for symmetry reasons that we are going to discuss later in this section, $g_{c}^2$ interaction terms are needed to provide a non-zero contributions to the bispectrum. \\
\noindent The propagators for ``plus'' and ``minus'' fields are
\bea\fl
\widehat{\delta B^{a,+}_{i}(\epr)\delta B^{b,+}_{j}(\eps)} = \tilde{\Pi}_{ij}^{ab}(\epr,\eps)\Theta(\epr-\eps)+\bar{\Pi}_{ij}^{ab}(\epr,\eps)\Theta(\eps-\epr),\\\fl
\widehat{\delta B^{a,+}_{i}(\epr)\delta B^{b,-}_{j}(\eps)} = \bar{\Pi}_{ij}^{ab}(\epr,\eps),\\\fl
\widehat{\delta B^{a,-}_{i}(\epr)\delta B^{b,+}_{j}(\eps)} = \tilde{\Pi}_{ij}^{ab}(\epr,\eps),\\\fl
\widehat{\delta B^{a,-}_{i}(\epr)\delta B^{b,-}_{j}(\eps)} = \bar{\Pi}_{ij}^{ab}(\epr,\eps)\Theta(\epr-\eps)+\tilde{\Pi}_{ij}^{ab}(\epr,\eps)\Theta(\eps-\epr),
\eea
or
\bea\fl
\tilde{\Pi}_{ij}^{ab}(\vec{k})\equiv T_{ij}^{even}(\hat{k})\tilde{P}^{ab}_{+}+i T_{ij}^{odd}(\hat{k})\tilde{P}^{ab}_{ij}+T_{ij}^{long}(\hat{k})\tilde{P}^{ab}_{ij}\\\fl
\bar{\Pi}_{ij}^{ab}(\vec{k})\equiv T_{ij}^{even}(\hat{k})\bar{P}^{ab}_{+}+i T_{ij}^{odd}(\hat{k})\bar{P}^{ab}_{ij}+T_{ij}^{long}(\hat{k})\bar{P}^{ab}_{ij}
\eea
in Fourier space. In the previous equations we set $\tilde{P}^{ab}_{\pm}\equiv (1/2)(\tilde{P}^{ab}_{R}\pm\tilde{P}^{ab}_{L})$, $\tilde{P}^{ab}_{R}=\delta_{ab}\delta B^{ab}_{R}(k,\eta^{*})\delta B^{*ab}_{R}(k,\eta)$ and $\bar{P}^{ab}_{\pm}=\left(\tilde{P}^{ab}_{\pm}\right)^{*}$ (similar definitions apply for $\tilde{P}^{ab}_{L}$ and $\tilde{P}^{ab}_{long}$).\\

\noindent We are now ready to show the computation of the following contributions to the bispectrum and trispectrum of $\zeta$
\bea\label{bi}\fl
\langle \zeta_{\vec{k_{1}}}\zeta_{\vec{k_{2}}}\zeta_{\vec{k_{3}}}\rangle \supset N^{i}_{a}N^{j}_{b}N^{k}_{c}\langle \delta A^{a}_{i}(\vec{k_{1}})\delta A^{b}_{j}(\vec{k_{2}})\delta A^{c}_{k}(\vec{k_{3}})\rangle, \\\label{tri}\fl
\langle\zeta_{\vec{k}_{1}}\zeta_{\vec{k}_{2}}\zeta_{\vec{k}_{3}}\zeta_{\vec{k}_{4}} \rangle \supset N_{i}^{a}N_{j}^{b}N_{k}^{c}N_{l}^{d}\langle \de A^{a}_{i}(\vec{k_{1}}) \de A^{b}_{j}(\vec{k_{2}}) \de A^{c}_{k}(\vec{k_{3}}) \de A^{d}_{l}(\vec{k_{4}})\rangle.
\eea\\

\noindent Eq.~(\ref{bi}) becomes
\bea\label{bi1}\fl
\langle \zeta_{\vec{k_{1}}}\zeta_{\vec{k_{2}}}\zeta_{\vec{k_{3}}}\rangle &\supset&N_{a}^{i}N_{b}^{j}N_{c}^{k}\frac{\delta^{(3)}(\vec{k_{1}}+\vec{k_{2}}+\vec{k_{3}})}{a^{3}(\eta^{*})}\Big[\int d \eta a^{4}(\eta) \tilde{\Pi}_{im}(\vec{k_{1}})\tilde{\Pi}^{l}_{j}(\vec{k_{2}})\tilde{\Pi}^{m}_{k}(\vec{k_{3}})\nonumber\\\fl&\times&\Big(g_{c}\ep^{abc}k_{1l}+g_{c}^2\ep^{eda}\ep^{ebc}B^{d}_{l}\Big)\Big]+perms.+c.c.
\eea
Even before performing the time integration, one realizes that, because of the antisymmetric properties of the Levi-Civita tensor, the $\sim g_{c}$ contribution on the r.h.s. of Eq.~(\ref{bi1}) is equal to zero once the sum over all the possible permutations has been performed. The vector bosons bispectrum is therefore proportional to $g_{c}^{2}$. The final result from (\ref{bi1}) has the following form

\bea\label{fire}\fl
\langle \zeta_{\vec{k_{1}}}\zeta_{\vec{k_{2}}}\zeta_{\vec{k_{3}}}\rangle \supset (2 \pi)^3 \delta^{(3)}(\vec{k}_{1}+\vec{k}_{2}+\vec{k}_{3})g_{c}^{2}H_{*}^{2}\sum_{n}F_{n}(k_{i},\eta^{*})I_{n}(\hat{k}_{i}\cdot\hat{k}_{j},\vec{A}_{i}\cdot\vec{A}_{j},\hat{k}_{i}\cdot\vec{A}_{j})
\eea
where $F_{n}$ are isotropic functions of time and of the moduli of the wave vectors ($i=1,2,3$) and $I_{n}$ are anisotropic coefficients. The sum in the previous equation is taken over all possible combinations of products of three polarization indices, i.e. $n\in (EEE, EEl, ElE,..., lll)$, where $E$ stands for ``even'', $l$ for ``longitudinal''. The complete expressions for the terms appearing in the sum are quite lengthy (see Sec.~$4.2$ of \cite{Bartolo:2009pa}). As an example, we report one of these terms
\bea\label{sample1}\fl
F_{lll}&=&-n^{6}(x^{*})\frac{1}{24 k^{6}k_{1}^{2}k_{2}^{2}k_{3}^{2}x^{*2}}\left[A_{EEE}+\left(B_{EEE}\cos x^{*}+C_{EEE}\sin x^{*}\right)E_{i}x^{*}\right]\\\label{sample2}\fl
I_{lll}&=&\ep^{aa'b'}\ep^{ac'e}\Big[\Big(\left(\hat{k}_{1}\cdot\vec{N}^{a'}\right)\left(\hat{k}_{3}\cdot\vec{N}^{b'}\right)\left(\hat{k}_{2}\cdot\vec{N}^{c'}\right)\left(\hat{k}_{1}\cdot\hat{k}_{2}\right)\left(\hat{k}_{3}\cdot\hat{A}^{e}\right)\nonumber\\\fl&-&\left(\hat{k}_{3}\cdot\vec{N}^{a'}\right)\left(\hat{k}_{2}\cdot\vec{N}^{b'}\right)\left(\hat{k}_{1}\cdot\vec{N}^{c'}\right)\left(\hat{k}_{1}\cdot\hat{k}_{2}\right)\left(\hat{k}_{3}\cdot\hat{A}^{e}\right)\Big)+(1\leftrightarrow 3)+(2\leftrightarrow 3) \Big]
\eea
where $A_{EEE}$, $B_{EEE}$ and $C_{EEE}$ are functions of $x^{*}$ and of the momenta $k_{i}\equiv |\vec{k}_{i}|$ (they are all reported in Appendix C of \cite{Bartolo:2009pa}), $E_{i}$ is the exponential-integral function and $i \leftrightarrow j$ means ``exchange $\hat{k}_{i}$ with $\hat{k}_{j}$''. As we will discuss in more details in 
Sec.~\ref{shapes}, one of the more interesting features is that the bispectrum and the trispectrum turn out to have an amplitude that is modulated 
by the preferred directions that break statistical isotropy.
\\

\noindent Let us now move to the trispectrum. Again, we count two different kinds of contributions, the first from $\sim g_{c}$ and the second from $\sim g_{c}^{2}$ interaction terms, respectively in $H_{int}^{(3)}$ and $H_{int}^{(4)}$. The former produce vector-exchange diagrams, the latter are represented by contact-interaction diagrams (see Fig.~2). Their analytic expressions are different, but they both have a structure similar to (\ref{fire}) 
\bea\label{fire1}\fl
\langle \zeta_{\vec{k_{1}}}\zeta_{\vec{k_{2}}}\zeta_{\vec{k_{3}}}\zeta_{\vec{k_{4}}}\rangle &\supset& (2 \pi)^3 \delta^{(3)}(\vec{k}_{1}+\vec{k}_{2}+\vec{k}_{3}+\vec{k}_{4}) g_{c}^{2}H_{*}^{2}\fl\\&\times&\sum_{n}G_{n}(k_{i},k_{\hat{12}},k_{\hat{14}},\eta^{*})L_{n}(\hat{k}_{i}\cdot\hat{k}_{j},\vec{A}_{i}\cdot\vec{A}_{j},\hat{k}_{i}\cdot\vec{A}_{j})\nonumber
\eea
where we define $k_{\hat{12}}\equiv |\vec{k}_{1}+\vec{k}_{2}|$ and $k_{\hat{14}}\equiv |\vec{k}_{1}+\vec{k}_{4}|$ (see Secs.~$5.2.1$ and $5.2.2$ of \cite{Bartolo:2009kg} for the explicit expressions of the functions $G_{n}$ and $L_{n}$).

\section{Amplitude of non-Gaussianity: $f_{NL}$ and $\tau_{NL}$}

In this review we use the following definitions for the non-Gaussianity amplitudes

\bea
\frac{6}{5}f_{NL}&=&\frac{B_{\zeta}(\vec{k}_{1},\vec{k}_{2},\vec{k}_{3})}{P^{iso}(k_{1})P^{iso}(k_{2})+perms.}\\
\tau_{NL}&=&\frac{2T_{\zeta}(\vec{k}_{1},\vec{k}_{2},\vec{k}_{3},\vec{k}_{4})}{P^{iso}(k_{1})P^{iso}(k_{2})P^{iso}(k_{\hat{14}})+23\,\,  {\rm perms.}}
\eea\\
\noindent The choice of normalizing the bispectrum and the trispectrum by the isotropic part of the power spectrum, instead of using its complete expression $P_{\zeta}$, is motivated by the fact that the latter would only introduce a correction to the previous equations proportional to the anisotropy parameter $g$, which is a small quantity.\\
The parameters $f_{NL}$ and $\tau_{NL}$ receive contributions both from scalar (``$s$'') and from vector (``$v$'') fields
\bea\label{uno}
f_{NL}=f_{NL}^{(s)}+f_{NL}^{(v)} ,\\
\tau_{NL}=\tau_{NL}^{(s)}+\tau_{NL}^{(v)} .
\eea
The latter can again be distinguished into Abelian ($A$) and non-Abelian ($NA$)
\bea
f_{NL}^{(v)}=f_{NL}^{(A)}+f_{NL}^{(NA)}  ,\\\label{due}
\tau_{NL}^{(v)}=\tau_{NL}^{(A_{1})}+\tau_{NL}^{(A_{2})}+\tau_{NL}^{({NA}_{1})}+\tau_{NL}^{({NA}_{2})}.
\eea 
The contribution $f_{NL}^{(A)}$ comes from Eq.~(\ref{b}), $f_{NL}^{(NA)}$ from (\ref{fire}), $\tau_{NL}^{(A_{1})}$ and $\tau_{NL}^{(A_{2})}$ from (\ref{t}), finally $\tau_{NL}^{({NA}_{1})}$ from (\ref{fire1}) and $\tau_{NL}^{({NA}_{2})}$ from the last line of (\ref{class4}).\\ 

\noindent Notice that, in order to keep the vector contributions manageable and simple in their structure, all gauge and vector indices have been purposely neglected at this point and so the angular functions appearing in the anisotropy coefficients have been left out of the final amplitude results. This is acceptable considering that these functions will in general introduce numerical corrections of order one. Nevertheless, it is important to keep in mind that the amplitudes also depend on the angular parameters of the theory. \\
We will now focus on the dependence of $f_{NL}$ and $\tau_{NL}$ from the non-angular parameters of the theory and quickly draw a comparison among the different contributions listed in Eqs.~(\ref{uno}) through (\ref{due}).\\ 

\begin{table}[t]\centering
\caption{Order of magnitude of $f_{NL}$ in different scenarios.\\}
\begin{tabular}{|c||c|c|c|}\hline 
 $$ & $f_{NL}^{s}$ & $f_{NL}^{A}$ & $f_{NL}^{NA}$ \\
\hline
\scriptsize{general case} &  \scriptsize{$\frac{1}{(1+\beta)^2}\frac{N_{\phi\phi}}{N_{\phi}^{2}}$} & \scriptsize{$\frac{\beta}{(1+\beta)^2}\frac{N_{AA}}{N_{\phi}^{2}}$} & \scriptsize{$\frac{\beta^2}{(1+\beta)^2}g_{c}^{2}\left(\frac{m}{H}\right)^2$} \\
\hline
\scriptsize{v.inflation} &  \scriptsize{$ \frac{\epsilon_{\phi}}{\left(1+\left(\frac{A}{m_{P}}\sqrt{\epsilon_{\phi}}\right)^2\right)^{2}}$} & \scriptsize{$\frac{\epsilon_{\phi}^2}{\left(1+\left(\frac{A}{m_{P}}\sqrt{\epsilon_{\phi}}\right)^2\right)^{2}} \left(\frac{A}{m_{P}}\right)^2 $} & \scriptsize{$\frac{\epsilon_{\phi}^2 g_{c}^{2}}{\left(1+\left(\frac{A}{m_{P}}\sqrt{\epsilon_{\phi}}\right)^2\right)^{2}} \left(\frac{A^2}{m_{P}H}\right)^2$} \\
\hline
\scriptsize{v.curvaton} & $\frac{\epsilon_{\phi}}{\left(1+\left(\frac{A m_{P}}{A_{tot}^2}\right)^{2}\epsilon_{\phi}r^2\right)^2}$ & $\frac{\epsilon_{\phi}^2 r^3}{\left(1+\left(\frac{A m_{P}}{A_{tot}^2}\right)^{2}\epsilon_{\phi}r^2\right)^2}\left(\frac{A m_{P}^{2}}{A_{tot}^{3}}\right)^2$ & $\frac{\epsilon_{\phi}^2 r^3 g_{c}^{2}}{\left(1+\left(\frac{A m_{P}}{A_{tot}^2}\right)^{2}\epsilon_{\phi} r^2\right)^2}\left(\frac{A^2 m_{P}^2}{A_{tot}^{3}H}\right)^2$ \\
\hline
\end{tabular}
\label{table1}
\end{table}

\begin{table}[t]\centering
\caption{Order of magnitude of the vector contributions to $\tau_{NL}$\\ in different scenarios.\\}
\begin{tabular}{|c||c|c|c|c|}\hline 
 $$ & \scriptsize{$\tau_{NL}^{{NA}_{1}}$} & \scriptsize{$\tau_{NL}^{{NA}_{2}}$} & \scriptsize{$\tau_{NL}^{A_{1}}$} & \scriptsize{$\tau_{NL}^{A_{2}}$}   \\
\hline 
\scriptsize{general case} & \scriptsize{$10^3\frac{\beta^2\epsilon g_{c}^{2}}{\left(1+\beta\right)^3}\left(\frac{m_{P}}{H}\right)^2$} & \scriptsize{$10^{-5}\frac{\beta^{3/2}\epsilon^{3/2}g_{c}^{2}}{\left(1+\beta\right)^3}\left(\frac{A}{H}\right)\left(\frac{m_{P}}{H}\right)m_{P}^2N_{AA}$} & \scriptsize{$\frac{\beta\epsilon^2}{\left(1+\beta\right)^3}m_{P}^4N_{AA}^2$} & \scriptsize{$\frac{\beta^{3/2}\epsilon^{3/2}}{\left(1+\beta\right)^3}m_{P}^3N_{AAA}$} \\
\hline
\scriptsize{v.inflation} & \scriptsize{same as above} & \scriptsize{$10^{-5}\frac{\beta^{3/2}\epsilon^{3/2}g_{c}^{2}}{\left(1+\beta\right)^3}\left(\frac{A}{H}\right)\left(\frac{m_{P}}{H}\right)$} & \scriptsize{$\frac{\beta\epsilon^2}{\left(1+\beta\right)^3}$} & $0$  \\
\hline
\scriptsize{v.curvaton} & \scriptsize{same as above} & \scriptsize{$10^{-5}\frac{r\beta^{3/2}\epsilon^{3/2}g_{c}^{2}}{\left(1+\beta\right)^3}\left(\frac{A}{H}\right)\left(\frac{m_{P}}{H}\right)\left(\frac{m_{P}}{A}\right)^2$} & \scriptsize{$\frac{r^2\beta\epsilon^2}{\left(1+\beta\right)^3}\left(\frac{m_{P}}{A}\right)^4$} & \scriptsize{$\frac{r\beta^{3/2}\epsilon^{3/2}}{\left(1+\beta\right)^3}\left(\frac{m_{P}}{A}\right)^3$} \\
\hline
\end{tabular}
\label{table1}
\end{table}

\noindent The expression of the number of e-foldings depends on the specific model and, in particular, on the mechanism of production of the fluctuations. Two possibilities have been described in Sec.~2. For vector inflation we have 
\bea
N_{a}^{i}=\frac{A^{a}_{i}}{2m_{P}^{2}},\quad\quad\quad\quad\quad\quad\quad\quad N_{ab}^{ij}=\frac{\delta_{ab}\delta^{ij}}{2m_{P}^{2}} 
\eea
(see Appendix B of \cite{Bartolo:2009pa} for their derivation). In the vector curvaton model the same quantities become \cite{Dimopoulos:2008yv,Bartolo:2009pa}
\bea
N_{a}^{i}=\frac{2}{3}r\frac{A_{i}^{a}}{\sum_{b}|\vec{A}^{b}|^2} ,\quad\quad\quad\quad\quad N_{ab}^{ij}=\frac{1}{3}r\frac{\delta_{ab}\delta^{ij}}{\sum_{c}|\vec{A}^{c}|^2}.
\eea
Neglecting tensor and gauge indices, the expressions above can be simplified as $N_{A}\simeq A/m_{P}^2$ and $N_{AA}\simeq 1/m_{P}^{2}$ in vector inflation, $N_{A}\simeq r/A$ and $N_{AA}\simeq r/A^{2}$ in the vector curvaton model. Also we have $N_{AAA}=0$ in vector inflation and $N_{AAA}\simeq r/A^3$ in vector curvaton.\\

\noindent We are now ready to provide the final expressions for the amplitudes: in Table~1 we list all the contributions to $f_{NL}$, Table~2 includes the vector contributions to $\tau_{NL}$, the scalar contributions being given by
\bea
\tau_{NL}^{(s)}=\frac{\epsilon_{\phi}}{\left(1+\beta\right)^3}+\frac{\epsilon^2_{\phi}}{\left(1+\beta\right)^3}.
\eea

\begin{table}[t]\centering
\caption{Order of magnitude of the ratios $f_{NL}^{v}/f_{NL}^{s}$ in different scenarios.\\}
\begin{tabular}{|c||c|c|}\hline 
 $$ & \scriptsize{$f_{NL}^{A}/f_{NL}^{s}$} & \scriptsize{$f_{NL}^{NA}/f_{NL}^{s}$}   \\
\hline 
\scriptsize{general case} & \scriptsize{$\beta\frac{N_{AA}}{N_{\phi\phi}}$} & \scriptsize{$\beta^2 g_{c}^{2}\left(\frac{m}{H}\right)^2\frac{N_{\phi}^{2}}{N_{\phi\phi}}$}  \\
\hline
\scriptsize{v.inflation} & \scriptsize{$\beta$ } & \scriptsize{$\frac{\beta^2 g_{c}^{2}}{\epsilon_{\phi}}\left(\frac{m_{P}}{H}\right)^2$}   \\
\hline
\scriptsize{v.curvaton} & \scriptsize{$\beta r\left(\frac{m_{P}}{A}\right)^2$} & \scriptsize{$\frac{\beta^2 g_{c}^{2}}{\epsilon_{\phi}r}\left(\frac{A}{H}\right)^2$}  \\
\hline
\end{tabular}
\label{table2}
\end{table}

\begin{table}[t]\centering
\caption{Order of magnitude of the ratios $\tau_{NL}^{v}/\tau_{NL}^{s}$ in different scenarios.\\}
\begin{tabular}{|c||c|c|c|c|}\hline 
 $$ & \scriptsize{$\tau_{NL}^{{NA}_{1}}/\tau_{NL}^{s}$} & \scriptsize{$\tau_{NL}^{{NA}_{2}}/\tau_{NL}^{s}$} & \scriptsize{$\tau_{NL}^{A_{1}}/\tau_{NL}^{s}$} & \scriptsize{$\tau_{NL}^{A_{2}}/\tau_{NL}^{s}$}   \\
\hline 
\scriptsize{general case} & \scriptsize{$10^3\beta^2 g_{c}^{2}\left(\frac{m_{P}}{H}\right)^2$} & \scriptsize{$10^{-5}\beta^{3/2}\epsilon^{1/2}g_{c}^{2}\left(\frac{A}{H}\right)\left(\frac{m_{P}}{H}\right)m_{P}^2N_{AA}$} & \scriptsize{$\beta\epsilon m_{P}^4N_{AA}^2$} & \scriptsize{$\beta^{3/2}\epsilon^{1/2}m_{P}^3N_{AAA}$} \\
\hline
\scriptsize{v.inflation} & \scriptsize{same as above} & \scriptsize{$10^{-5}\beta^{3/2}\epsilon^{1/2}g_{c}^{2}\left(\frac{A}{H}\right)\left(\frac{m_{P}}{H}\right)$} & \scriptsize{$\beta\epsilon$} & $0$  \\
\hline
\scriptsize{v.curvaton} & \scriptsize{same as above} & \scriptsize{$10^{-5}r\beta^{3/2}\epsilon^{1/2}g_{c}^{2}\left(\frac{A}{H}\right)\left(\frac{m_{P}}{H}\right)\left(\frac{m_{P}}{A}\right)^2$} & \scriptsize{$r^2\beta\epsilon\left(\frac{m_{P}}{A}\right)^4$} & \scriptsize{$r\beta^{3/2}\epsilon^{1/2}\left(\frac{m_{P}}{A}\right)^3$} \\
\hline
\end{tabular}
\label{table2}
\end{table}

\noindent In the expressions appearing in the tables, numerical coefficients of order one have not been reported. Also, $m$ is by definition equal to $m_{P}$ in vector inflation and to $A/\sqrt{r}$ in the vector curvaton model; $N_{\phi}\simeq (m_{P}\sqrt{\epsilon_{\phi}})^{-1}$ and $N_{\phi\phi}\simeq m_{P}^{-2}$, with $\epsilon_{\phi}\equiv(\dot{\phi}^2)/(2 m_{P}^{2}H^{2})$.\\

\noindent The quantities involved in the amplitude expressions are $g$, $\beta$, $r$, $\epsilon_{\phi}$, $g_{c}$, $m_{P}/H$, $A/m_{P}$ and $A/H$. We already know that $g$ and $\beta$ are to be considered smaller than one (see discussion after Eq.~(\ref{pdisc})). Similarly, as mentioned after Eq.~(\ref{indirect}), $r$ has to remain small at least until inflation ends so as to attain an ``almost isotropic'' expansion. The slow-roll parameter $\epsilon_{\phi}$ and the $SU(2)$ coupling $g_{c}$ are small respectively to allow the inflaton to slowly roll down its potential and for perturbation theory to be valid. The ratio $m_{P}/H$ is of order $10^{5}$ (assuming $\epsilon_{\phi}\sim 10^{-1}$). Finally, $A/m_{P}$ and $A/H$ have no stringent bounds. A reasonable choice could be to assume that the expectation value of the gauge fields is no larger than the Planck mass, i.e. $A/m_{P}\leq 1$. As to the $A/H$ ratio, different possibilities are allowed, including the one where it is of order one (see Sec.~6 and Appendix A of \cite{Bartolo:2009pa} for a discussion on this).\\

\noindent Let us now compare the different amplitude contributions. The ratios between scalar and vector contributions are shown in Table~3 for the bispectrum and Table~4 for the trispectrum. We can observe that the dominance of a given contribution w.r.t. another one very much depends on the selected region of parameter space. It turns out that it is allowed for the vector contributions to be larger than the scalar ones and also for the non-Abelian contributions to be larger than the Abelian ones. This is discussed more in details in Sec.~6 of \cite{Bartolo:2009pa}. An interesting point is, for instance, the following: ignoring tensor and gauge indices, the ratio $g_{c}A/H$, that appears in many of the Tables entries, is a quantity smaller than one; if we consider the different configurations identified by gauge and vector indices, we realize that this is not always true, in fact the value of this ratio can be $\gg 1$ in some configurations. \\
Finally, it is interesting to compare bispectrum and trispectrum amplitudes (see table~5). Again, it is allowed for the ratios appearing in Table~5 to be either large or small, depending on the specific location within the parameter space of the theory. For instance, the combination of a small bispectrum with a large trispectrum is permitted. The latter is an interesting possibility: if the bispectrum was observably small, we could still hope the information about non-Gaussianity to be accessible thanks to the trispectrum.\\ 
Another interesting feature of this model is that the bispectrum and the trispectrum depend on the same set of quantities. If these correlation functions were independently known, that information could then be used to test the theory and place some bounds on its parameters.

\begin{table}[t]\centering
\caption{Order of magnitude of the ratios $\tau_{NL}^{v}/\left(f_{NL}^{NA}\right)^2$ in different scenarios.\\}
\begin{tabular}{|c||c|c|c|c|}\hline 
 $$ & \scriptsize{$\tau_{NL}^{{NA}_{1}}/\left(f_{NL}^{NA}\right)^2$} & \scriptsize{$\tau_{NL}^{{NA}_{2}}/\left(f_{NL}^{NA}\right)^2$} & \scriptsize{$\tau_{NL}^{A_{1}}/\left(f_{NL}^{NA}\right)^2$} & \scriptsize{$\tau_{NL}^{A_{2}}/\left(f_{NL}^{NA}\right)^2$}   \\
\hline 
\scriptsize{v.i.} & \scriptsize{$10^9\frac{\epsilon\left(1+\beta\right)}{g_{c}^{2}\beta^2}\left(\frac{H}{m_{P}}\right)^2$} & \scriptsize{$10\frac{\epsilon^{3/2}\left(1+\beta\right)}{\beta^{5/2}g_{c}^{2}}\left(\frac{A}{H}\right)\left(\frac{H}{m_{P}}\right)^3$} & \scriptsize{$10^6\frac{\epsilon^2\left(1+\beta\right)}{\beta^3g_{c}^{4}}\left(\frac{H}{m_{P}}\right)^4$} & $0$  \\
\hline
\scriptsize{v.c.} & \scriptsize{$10^9\frac{r^2\epsilon\left(1+\beta\right)}{g_{c}^{2}\beta^2}\frac{m_{P}^{2}}{A^{2}}\frac{H^{2}}{A^{2}}$} & \scriptsize{$10\frac{r^5\epsilon^{3/2}\left(1+\beta\right)}{\beta^{5/2}g_{c}^{2}}\frac{H^{3}}{A^{3}}\frac{m_{P}}{H}\frac{m_{P}^{2}}{A^{2}}$} & \scriptsize{$10^6\frac{r^6\epsilon^2\left(1+\beta\right)}{\beta^3 g_{c}^{4}}\left(\frac{m_{P}}{A}\right)^4\left(\frac{H}{A}\right)^4$} & \scriptsize{$10^6\frac{r^3\epsilon^{3/2}\left(1+\beta\right)}{g_{c}^{2}\beta^{5/2}}\frac{m_{P}^{3}}{A^{3}}\frac{H^{4}}{A^{4}}$} \\
\hline
\end{tabular}
\label{table3}
\end{table}

\section{Shape of non-Gaussianity and statistical anisotropy features}
\label{shapes}

Studying the shape of non-Gaussianity means understanding the features of momentum dependence of the bispectrum and higher order correlators. If they also depend on variables other than momenta, it is important to determine how these other variables affect the profiles for any given momentum set-up. This is the case as far as the bispectrum and the trispectrum of the gauge fields are concerned, given the fact that they are functions, besides of momenta, also of a large set of angular variables (see Eqs.~(\ref{fire}) and (\ref{fire1})).

\subsection{Momentum dependence of the bispectrum and trispectrum profiles}

\noindent We show the study of the momentum dependence of the $F_{n}$ and $G_{n}$ functions in Eqs.~(\ref{fire}) and (\ref{fire1}) first and then analyze the angular variables dependence of the spectra, once the momenta have been fixed in a given configuration. A natural choice would be to consider the configuration where the correlators are maximized. \\
The maxima can be easily determined for the bispectrum by plotting the isotropic functions $F_n$ and $G_n$ in terms of two of their momenta. These plots are provided in Fig.~3, where the variables are $x_{2}\equiv k_{2}/k_{1}$ and $x_{3}\equiv k_{3}/k_{1}$. Each one of the plots corresponds to a single isotropic functions of the sum in Eq.~(\ref{fire}). It is apparent that the maxima are mostly located in the in the so-called local region, i.e. for $k_{1}\sim k_{2}\gg k_{3}$; three out of the eight graphs do not have their peaks in this configuration but, at the same time, they show negligible amplitudes compared to the ``local'' peaked graphs.

\begin{figure}\centering
 \includegraphics[width=0.4\textwidth]{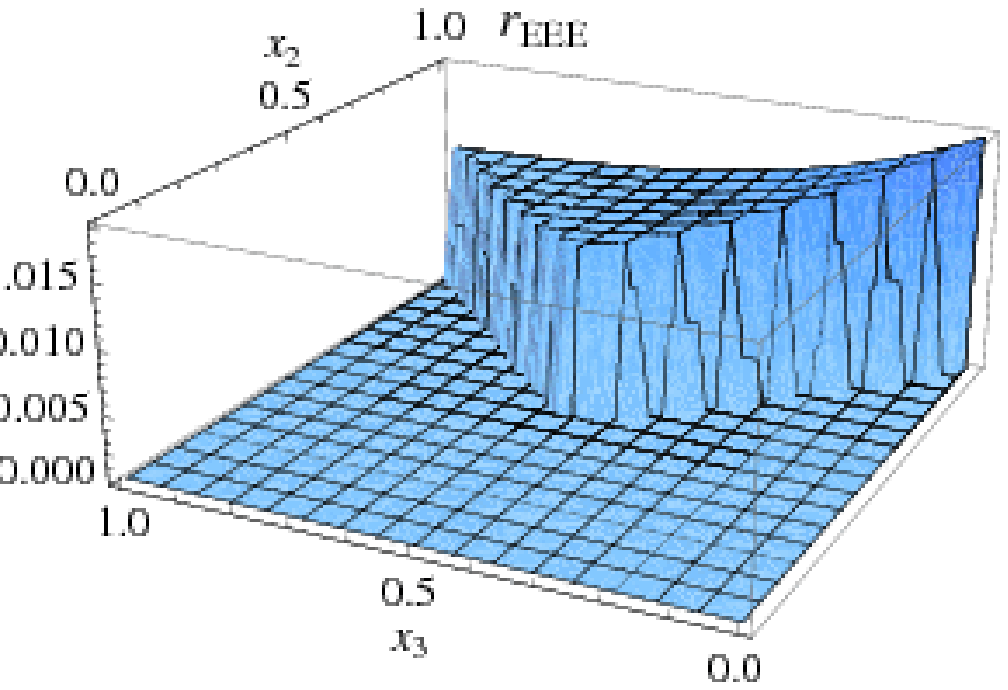}
\hspace{0.1\textwidth}
 \includegraphics[width=0.4\textwidth]{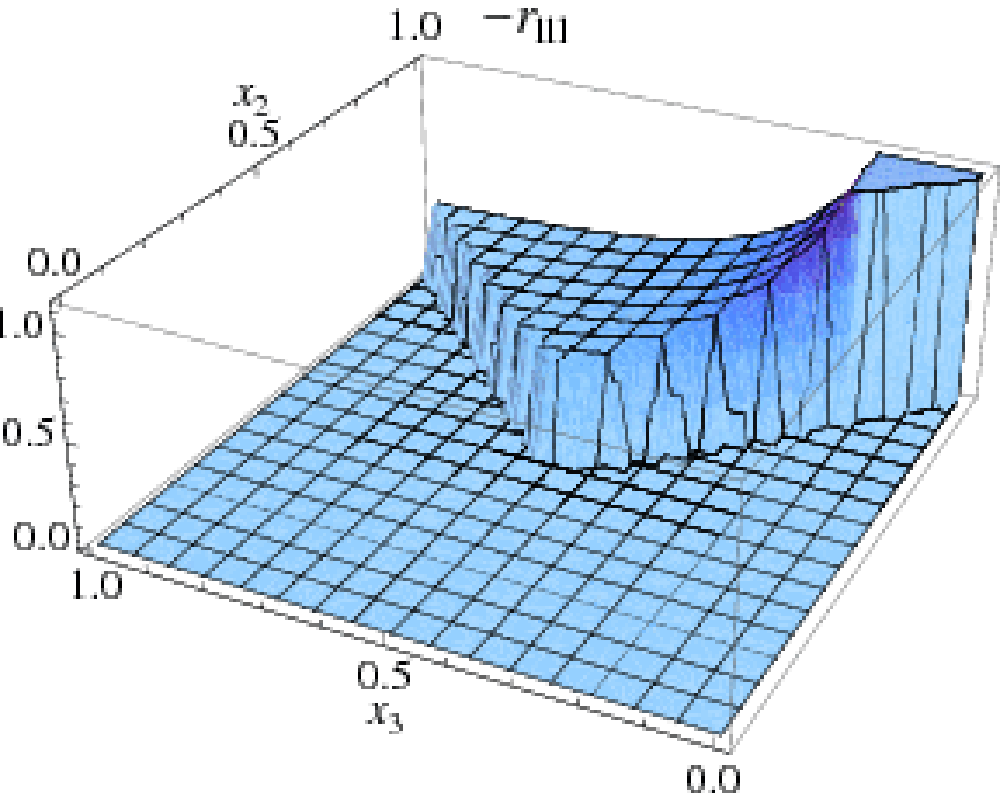}
\vspace{0.02\textwidth}
 \includegraphics[width=0.4\textwidth]{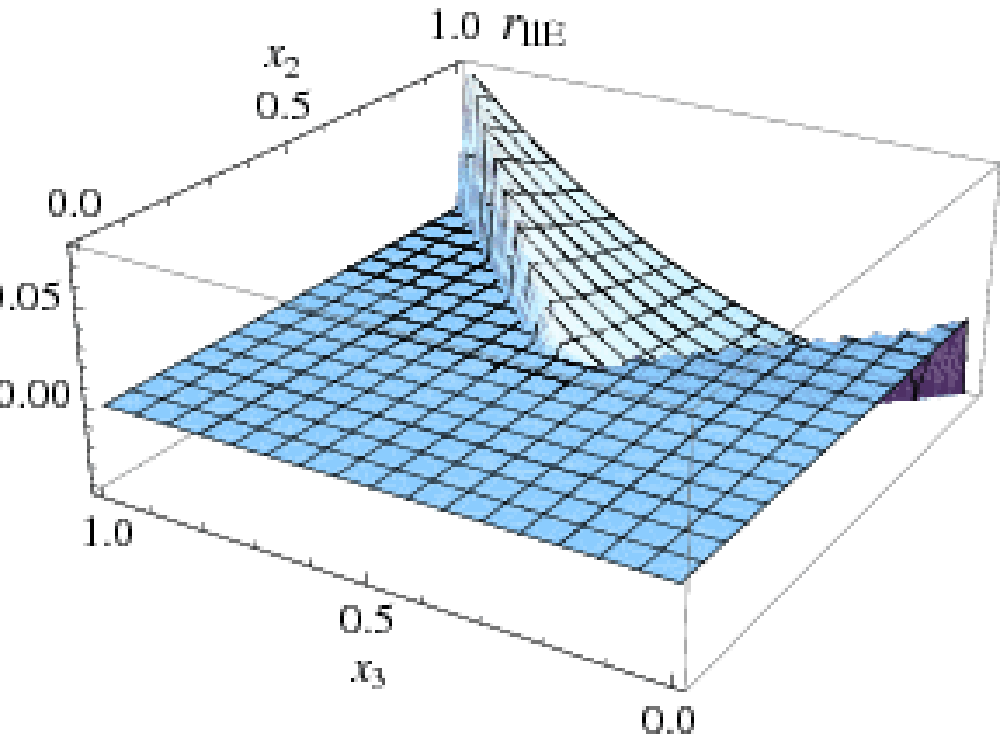}
\hspace{0.1\textwidth}
 \includegraphics[width=0.4\textwidth]{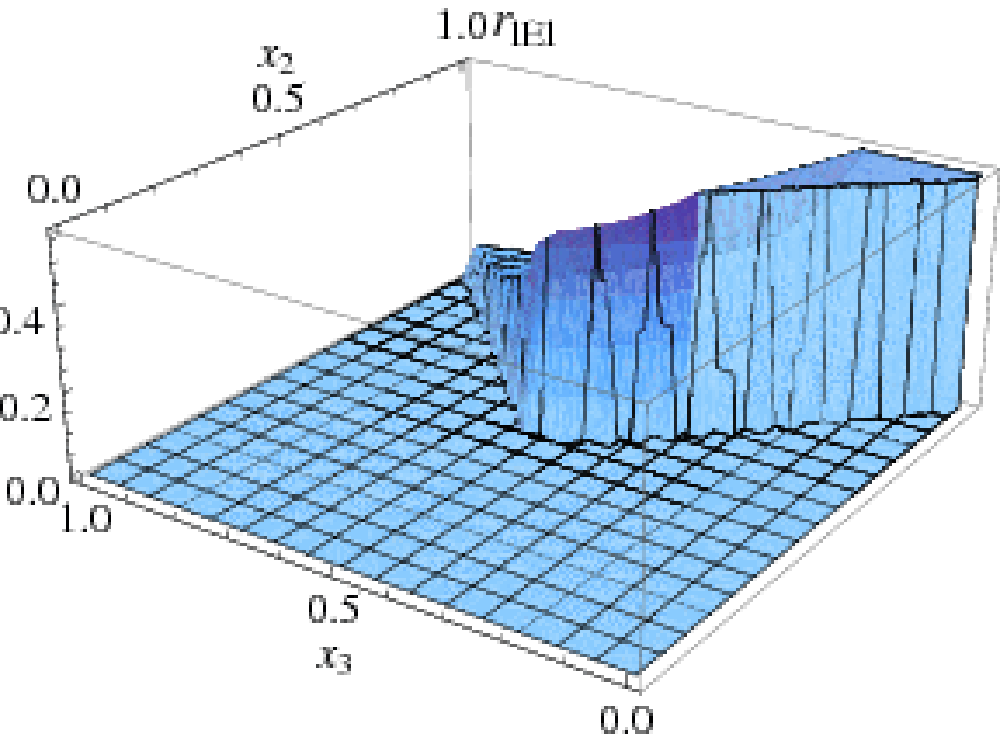}
\vspace{0.02\textwidth}
 \includegraphics[width=0.4\textwidth]{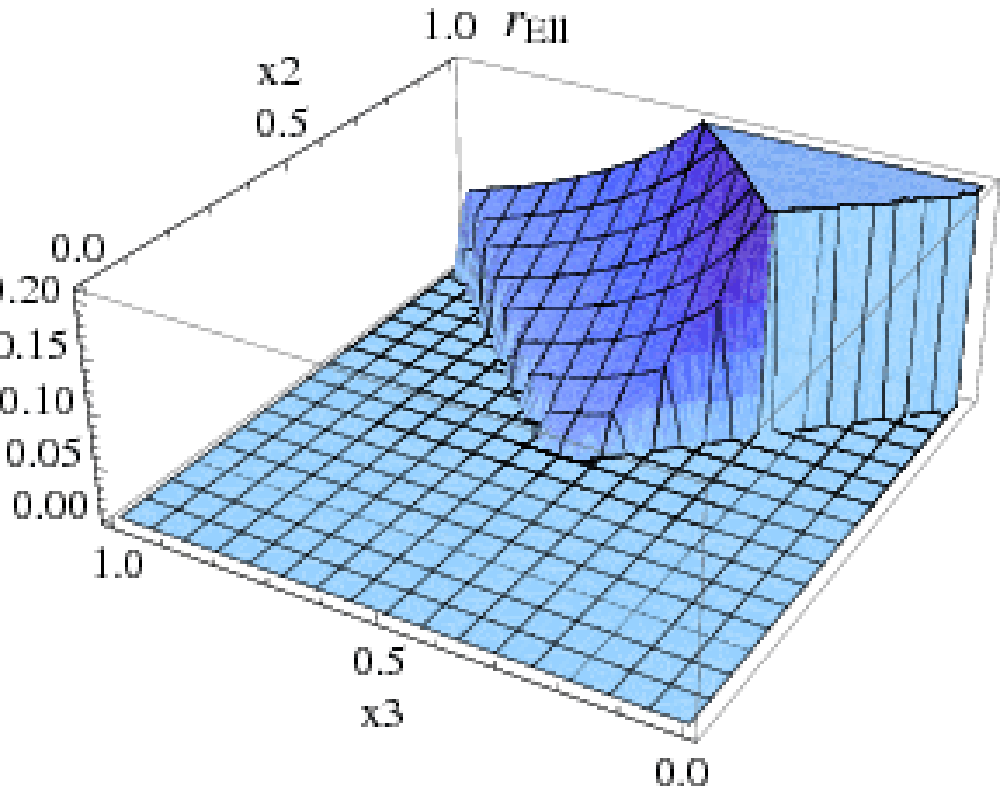}
\hspace{0.1\textwidth}
 \includegraphics[width=0.4\textwidth]{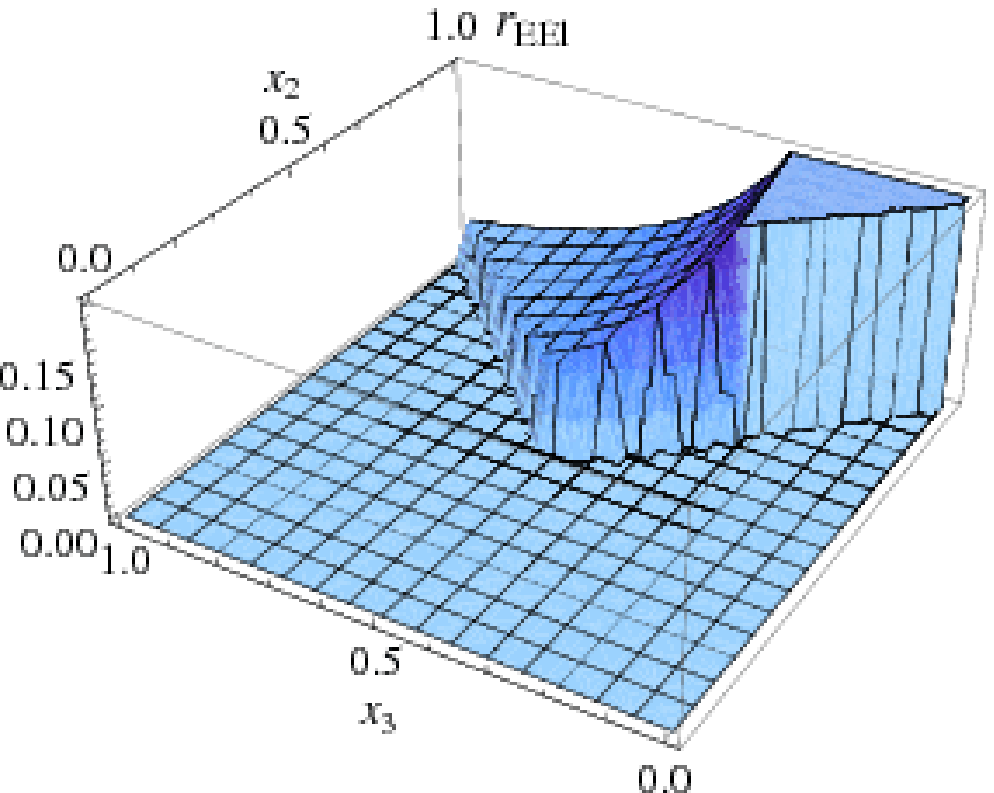}
\vspace{0.02\textwidth}
 \includegraphics[width=0.4\textwidth]{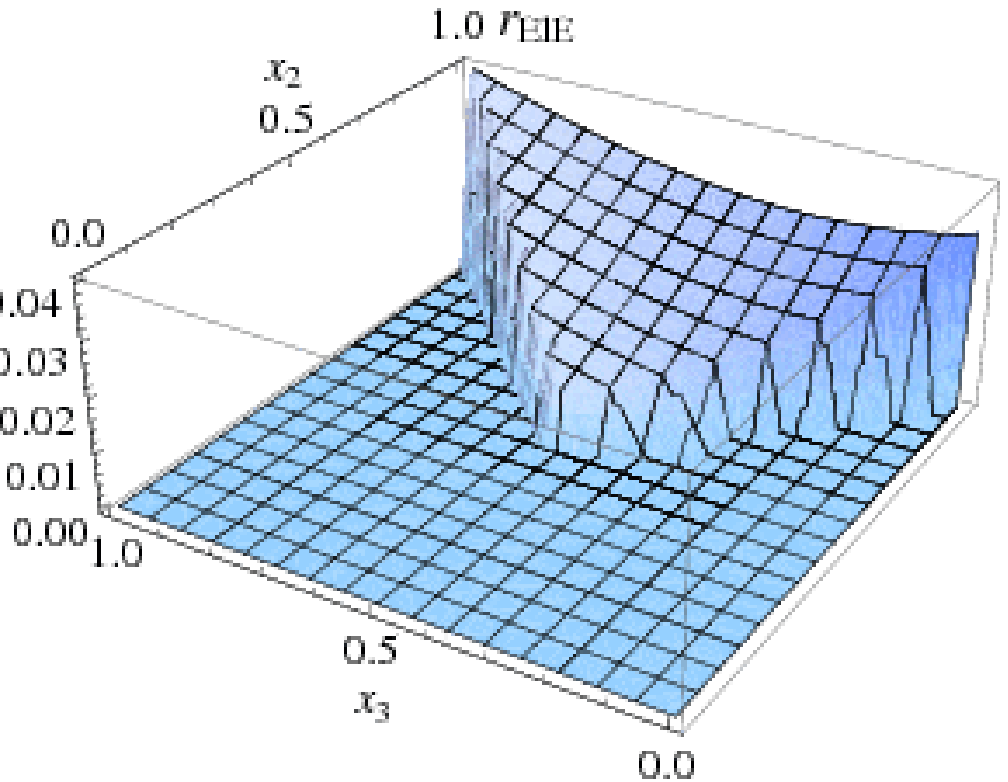}
\hspace{0.1\textwidth}
 \includegraphics[width=0.4\textwidth]{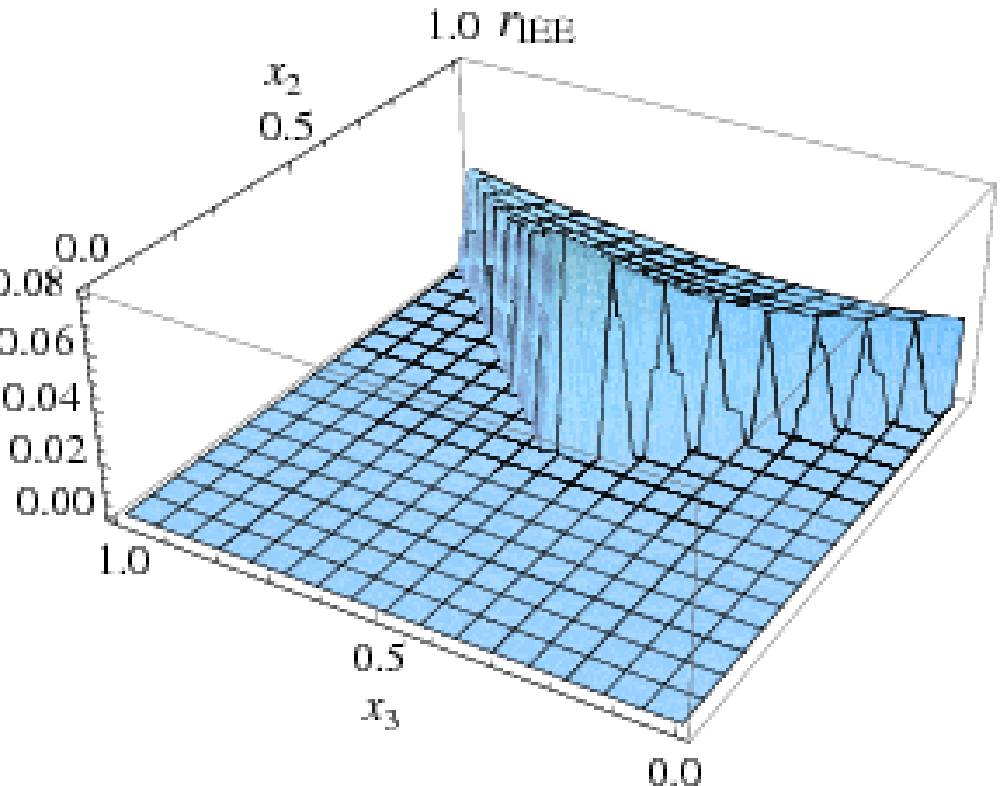}
\caption{ \label{Fig1}Plot of $r_{n}\equiv \Theta(x_{2}-x_
{3})\Theta(x_{3}-1+x_{2}) x_{2}^{2}x^{2}_{3}R_{n}(x_{2},x_{3})$, where\\we define $R_{n}=k_{1}^{6}F_{n}$. The Heaviside step functions $\Theta$ help
restricting\\the plot domain to the region $(x_{2},x_{3})$ that is allowed
for the triangle\\$\vec{k}_{1}+\vec{k}_{2}+\vec{k}_{3}=0$ (in particular, we
set $x_{3}<x_{2}$). We also set $x^{*}=1$.}
\end{figure}

\noindent The situation is much more complex for the trispectrum, being the number of momentum variables larger than three ($k_{1}$, $k_{2}$, $k_{3}$, $k_{4}$, $k_{\hat{12}}$ and $k_{\hat{14}}$). The momentum dependence of the isotropic functions can be studied by selecting different configurations for the tetrahedron made up by the four momentum vectors, in such a way as to narrow the number of independent momentum variables down to two. A list of possible configurations was presented in \cite{Chen:2009bc}. We consider two of them, the ``equilateral'' and the ``specialized planar''.

\begin{figure}\centering
\includegraphics[width=0.4\textwidth]{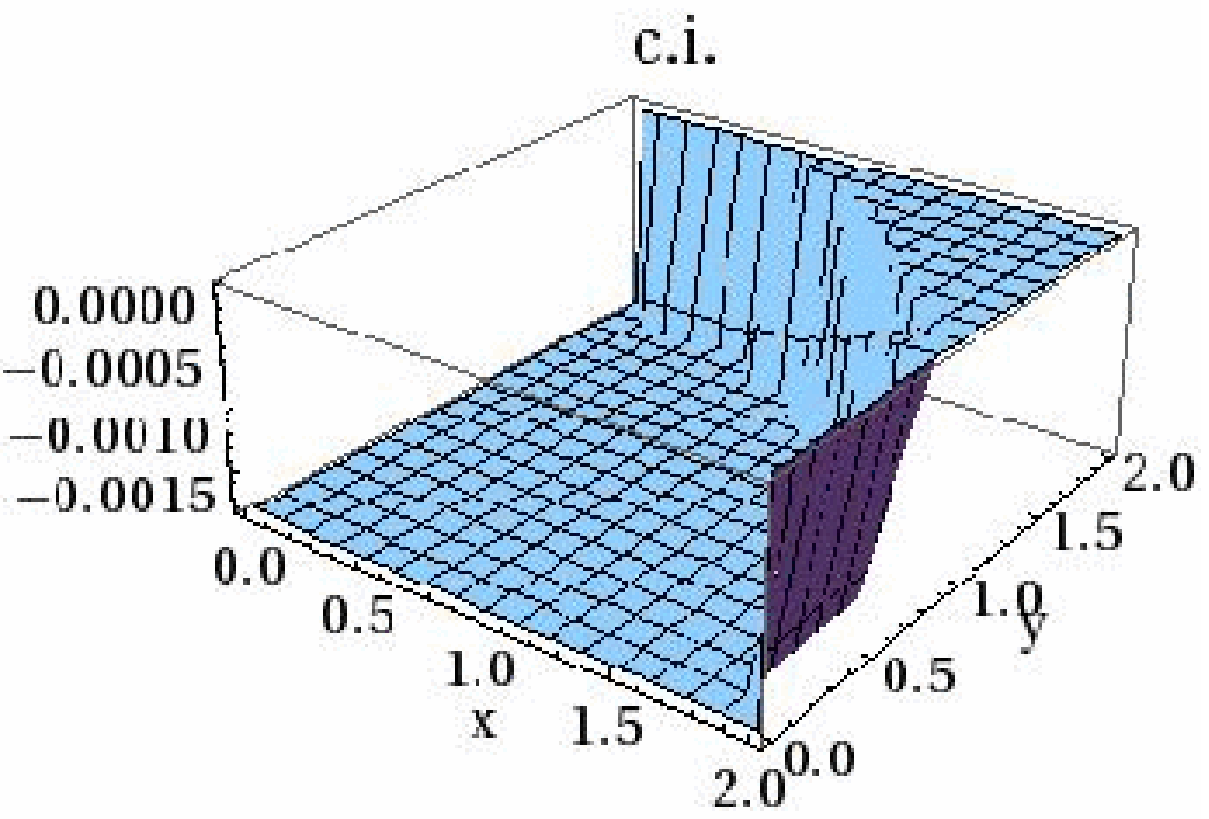}
\hspace{0.1\textwidth}
 \includegraphics[width=0.4\textwidth]{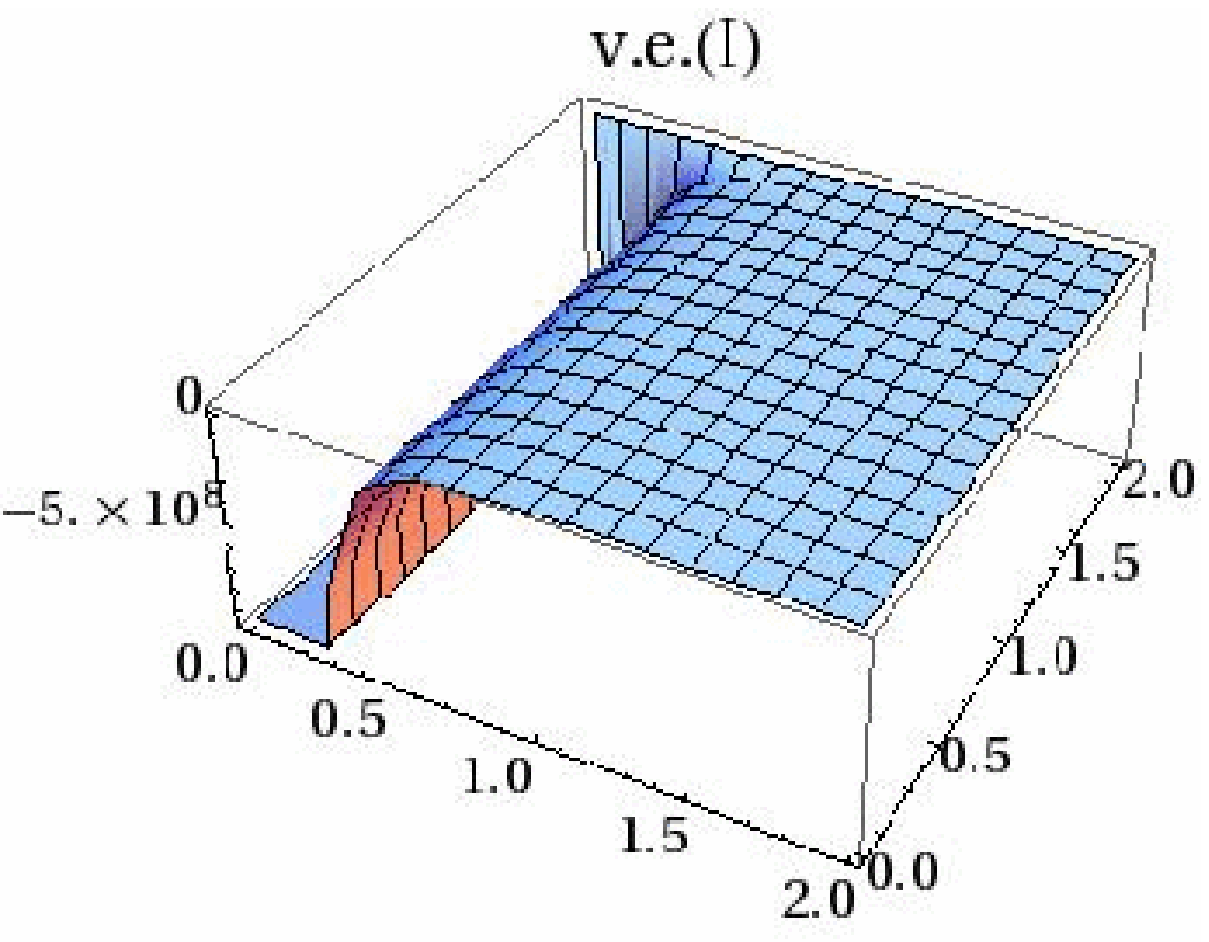}
\vspace{0.02\textwidth}
 \includegraphics[width=0.4\textwidth]{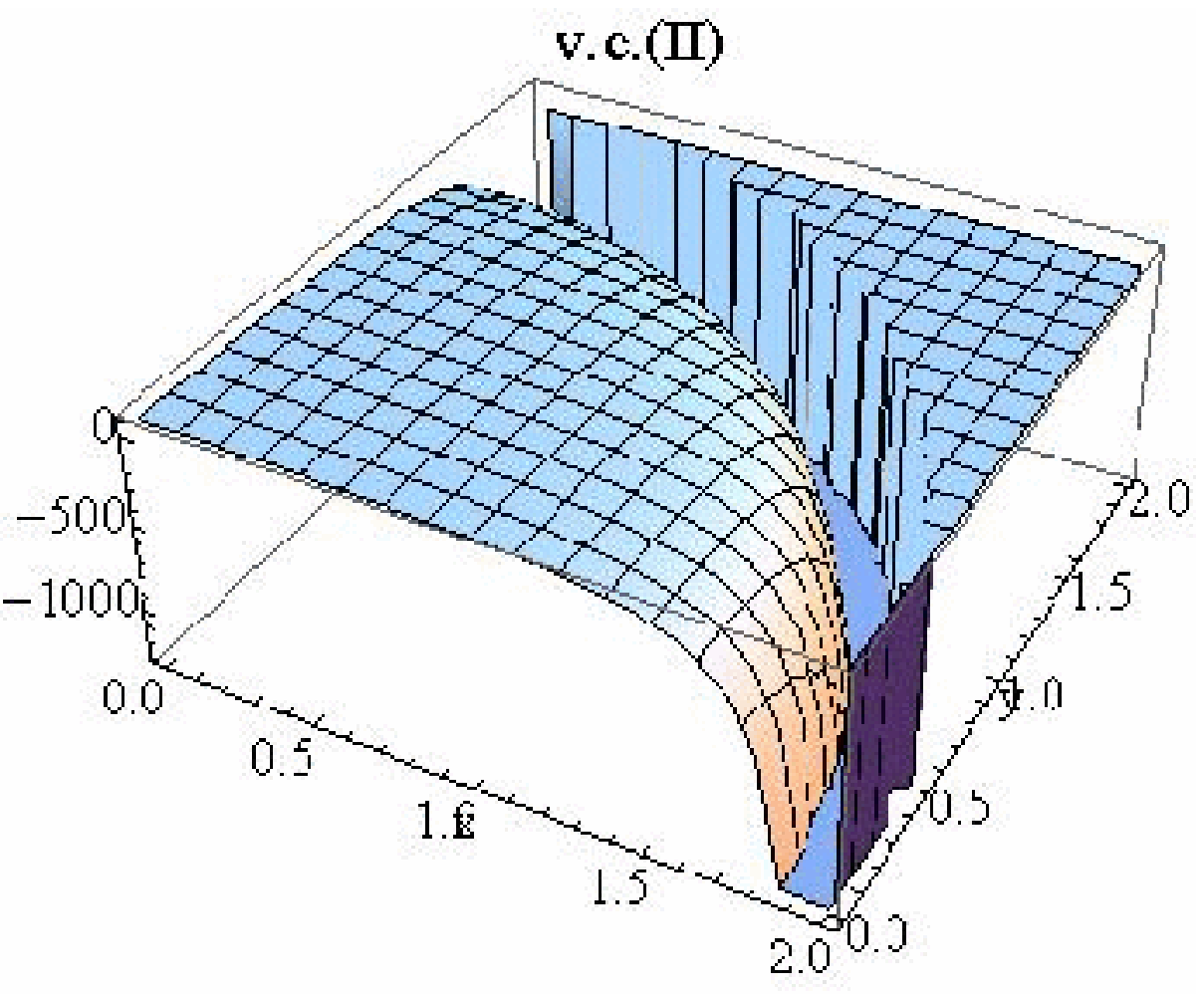}
\hspace{0.1\textwidth} \includegraphics[width=0.4\textwidth]{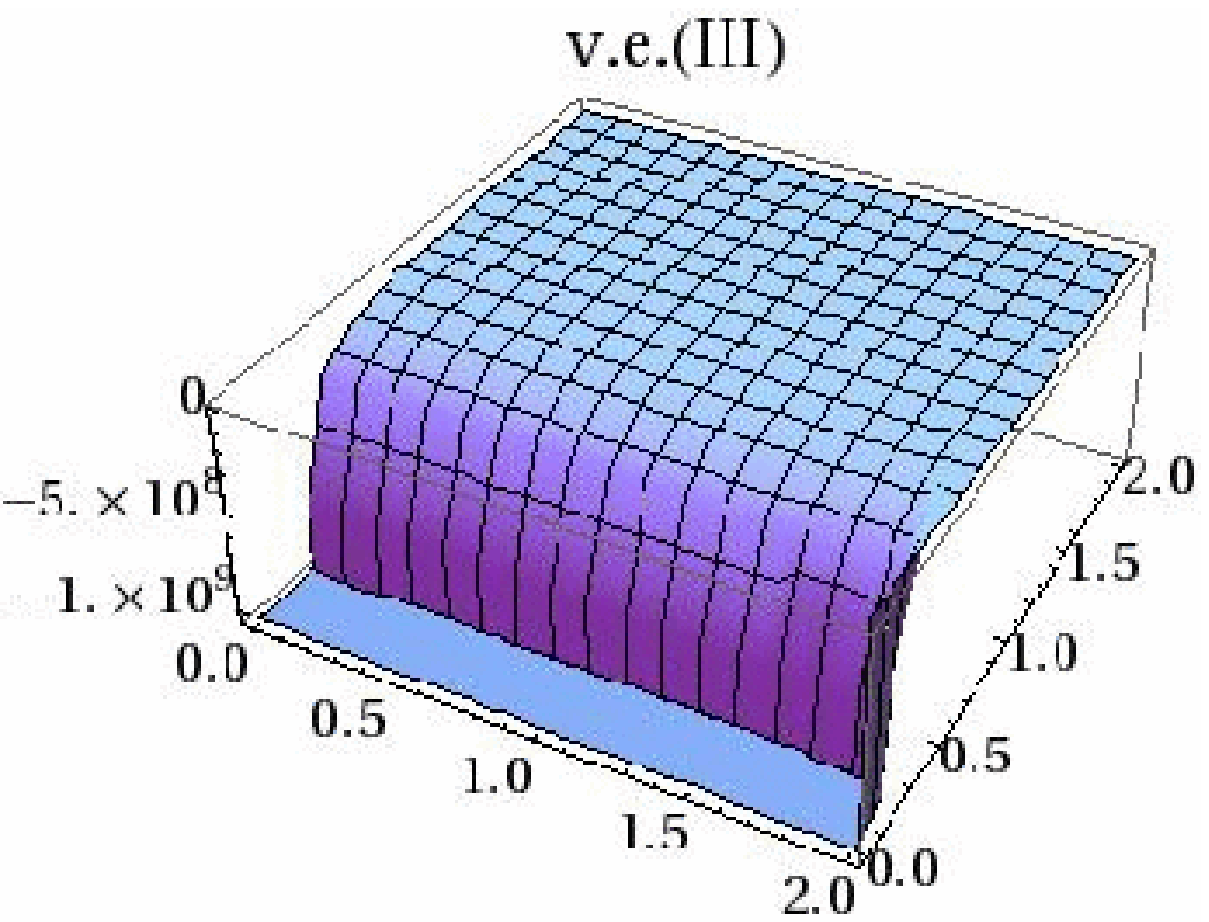}
\caption{ \label{Fig2} Plots of the isotropic functions appearing in the vector fields trispectrum (from Eq.~(\ref{fire1})): $c.i.$ is the contribution from contact-interaction diagrams, $v.e.(I)$, $v.e.(II)$ and $v.e.(III)$ are the contributions from the vector-exchange diagrams. The equilateral configuration has been considered in this figure.}
\end{figure}

\begin{figure}[t]\centering
\includegraphics[width=0.4\textwidth]{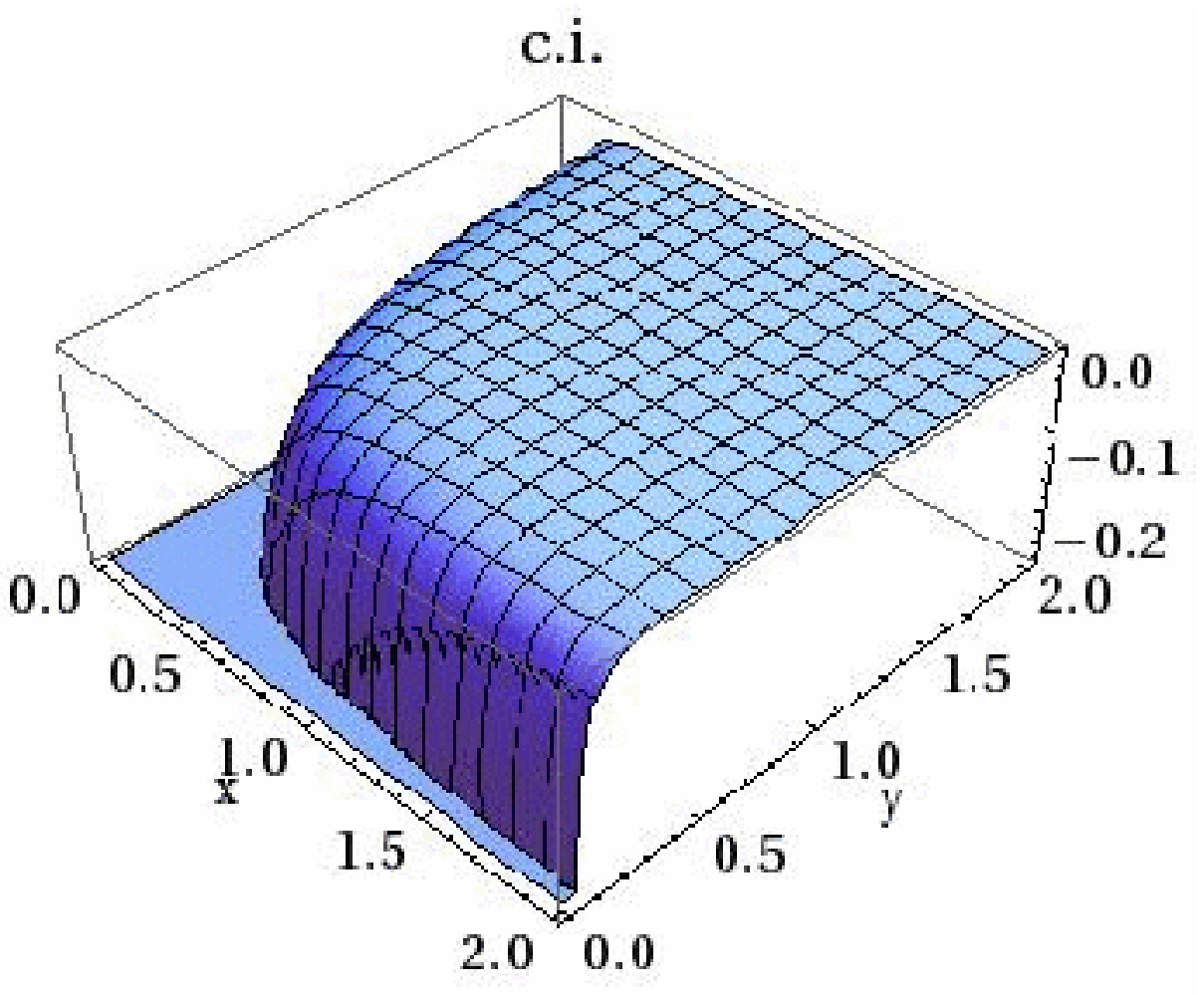}
\hspace{0.1\textwidth}
 \includegraphics[width=0.4\textwidth]{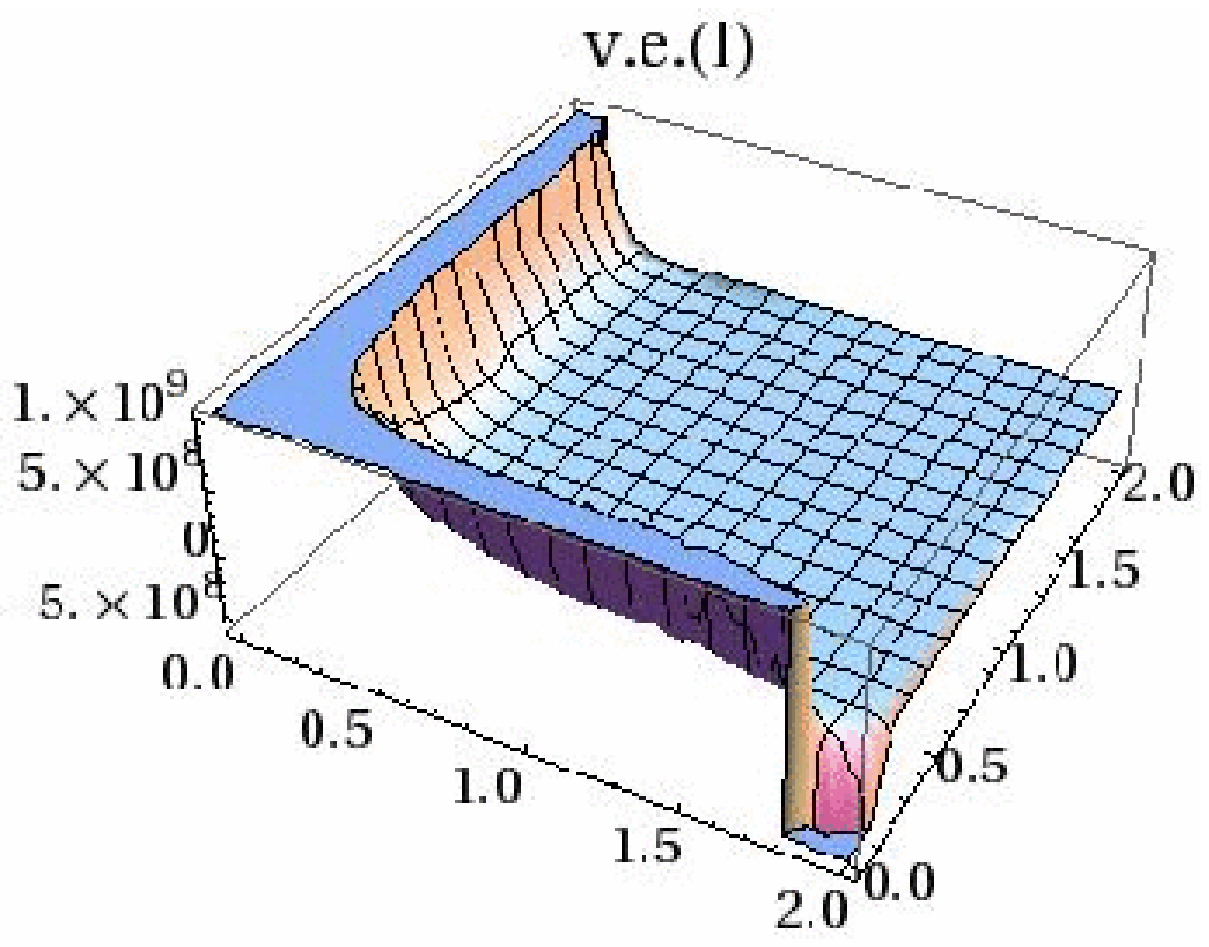}
\vspace{0.02\textwidth}
 \includegraphics[width=0.4\textwidth]{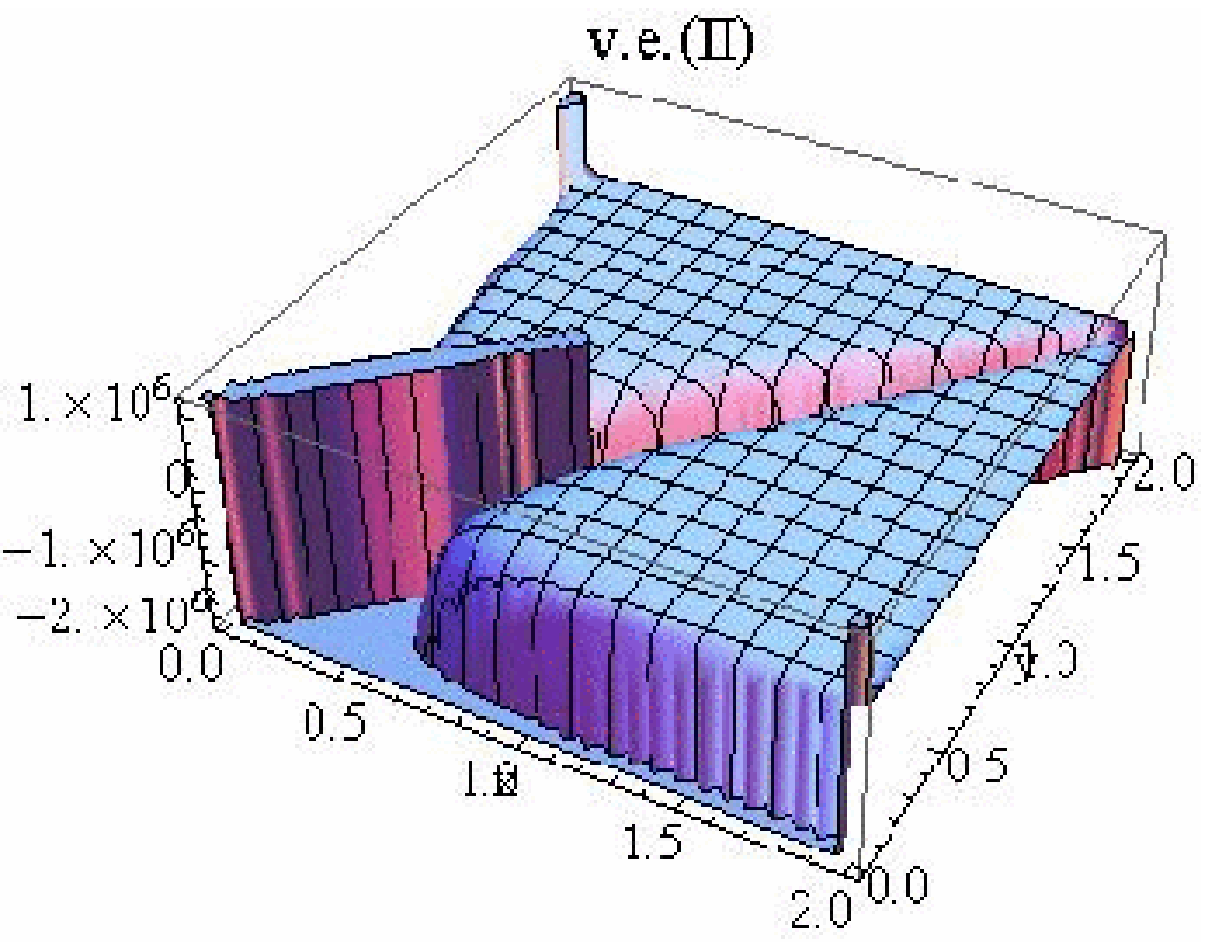}
\hspace{0.1\textwidth}
 \includegraphics[width=0.4\textwidth]{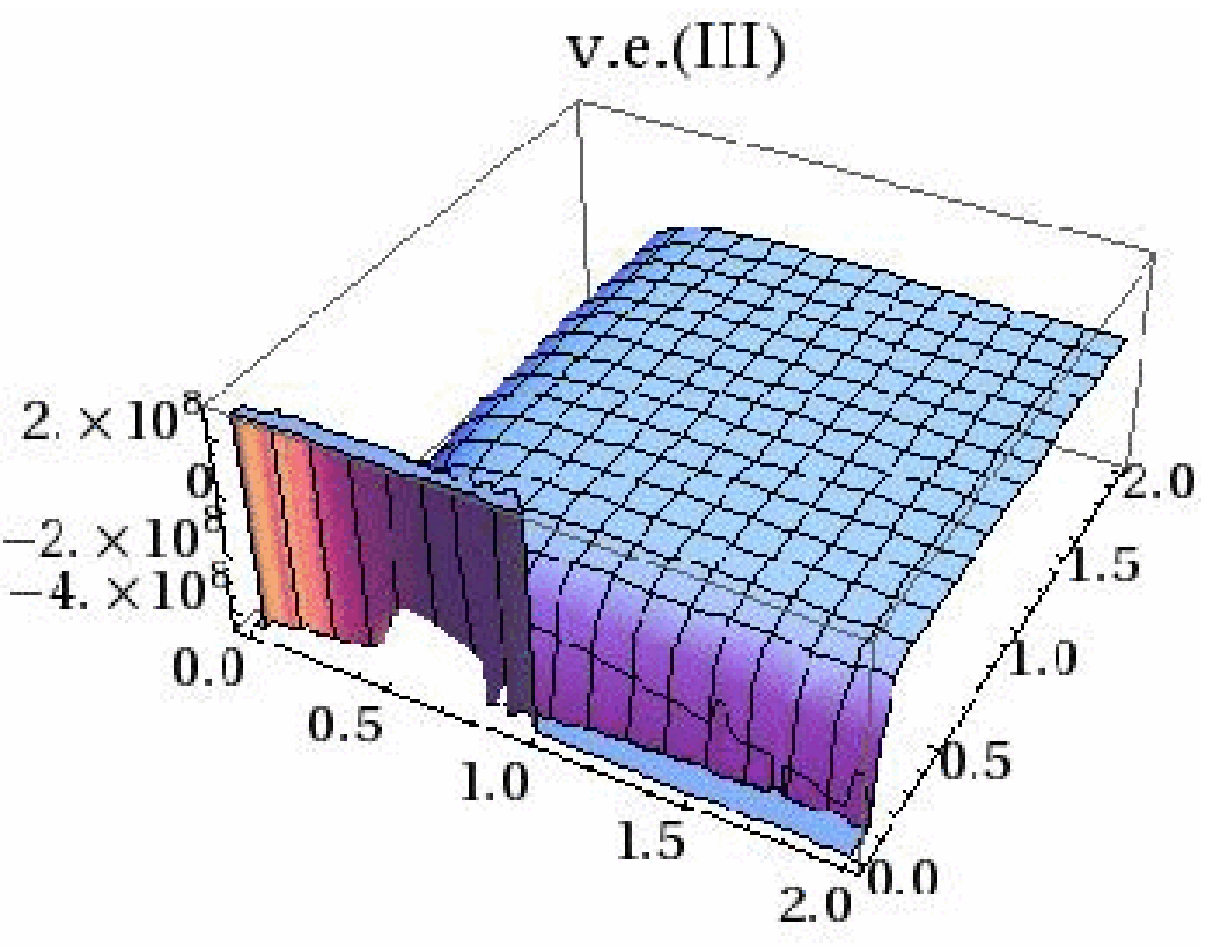}
\caption{ \label{Fig3} Plots of the contact interaction and of the vector-exchange contributions in the specialized planar configuration (plus sign).}
\end{figure}

\begin{figure}\centering
\includegraphics[width=0.4\textwidth]{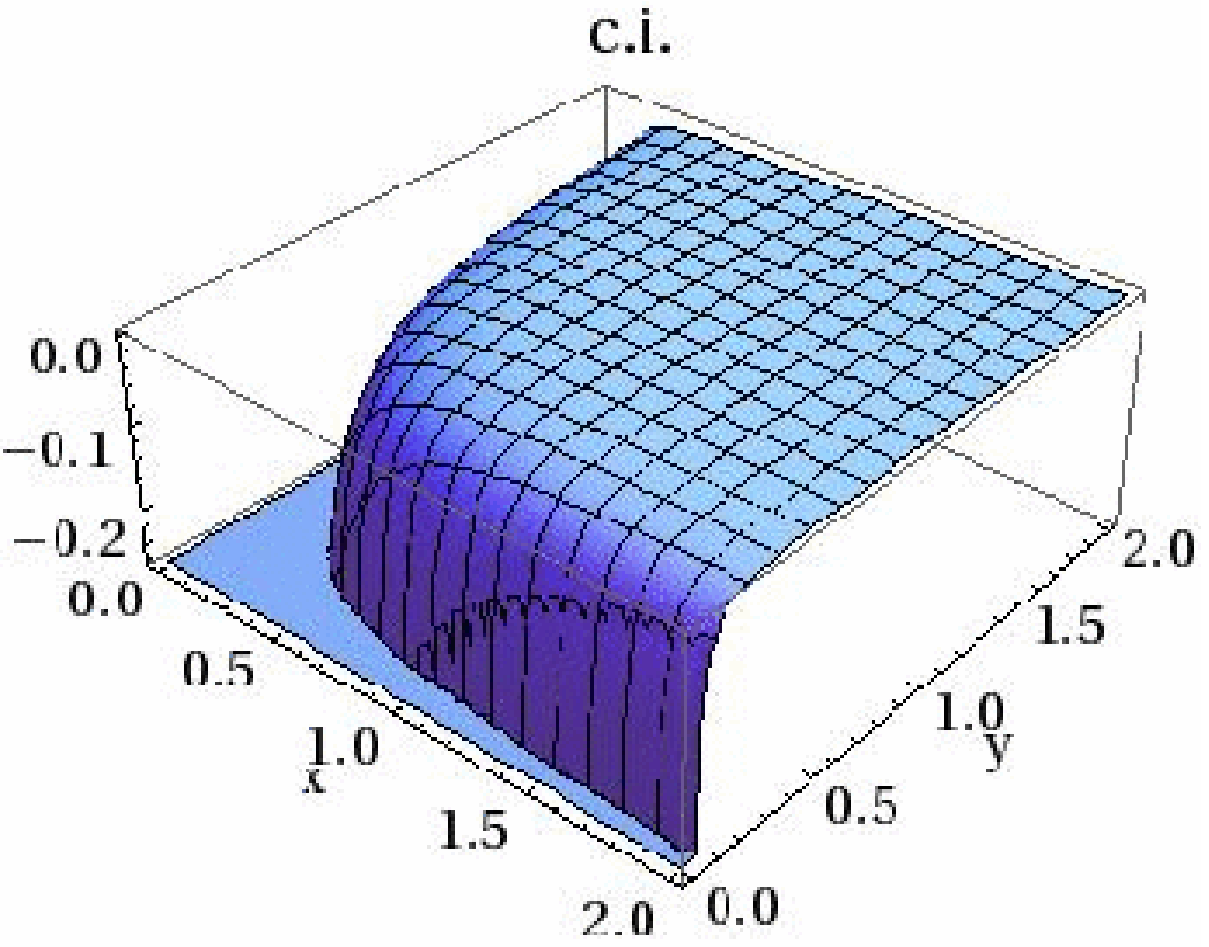}
\hspace{0.1\textwidth}
 \includegraphics[width=0.4\textwidth]{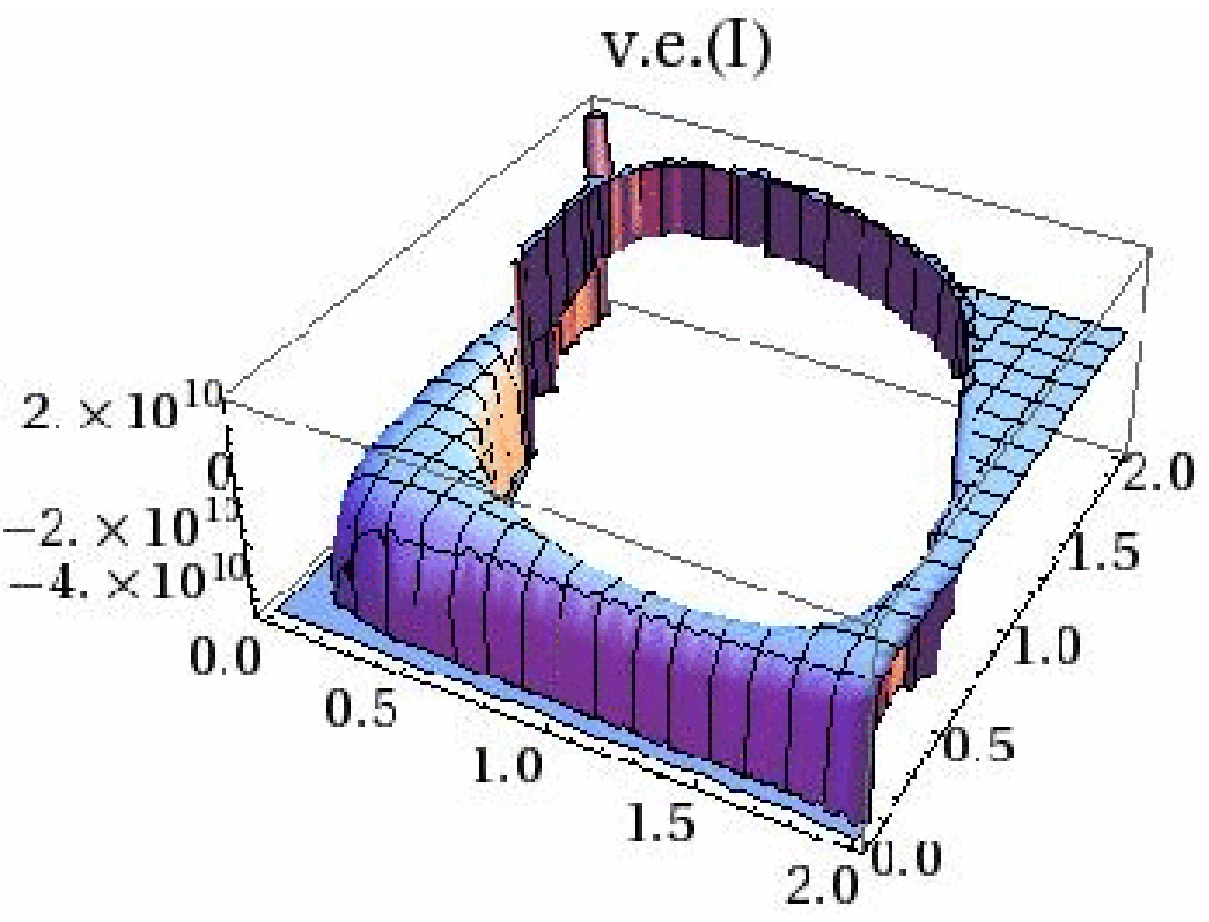}
\vspace{0.02\textwidth}
 \includegraphics[width=0.4\textwidth]{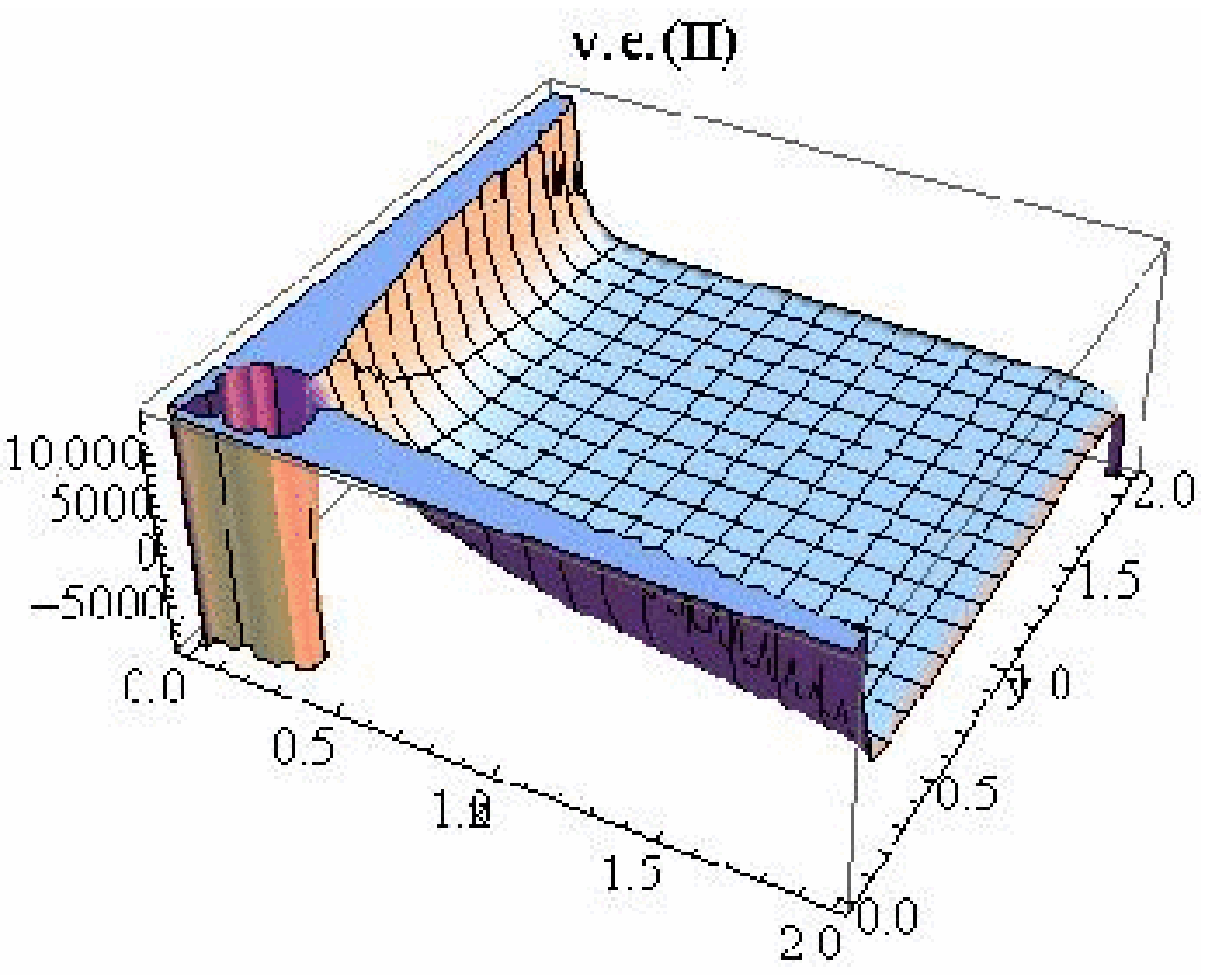}
\hspace{0.1\textwidth}
 \includegraphics[width=0.4\textwidth]{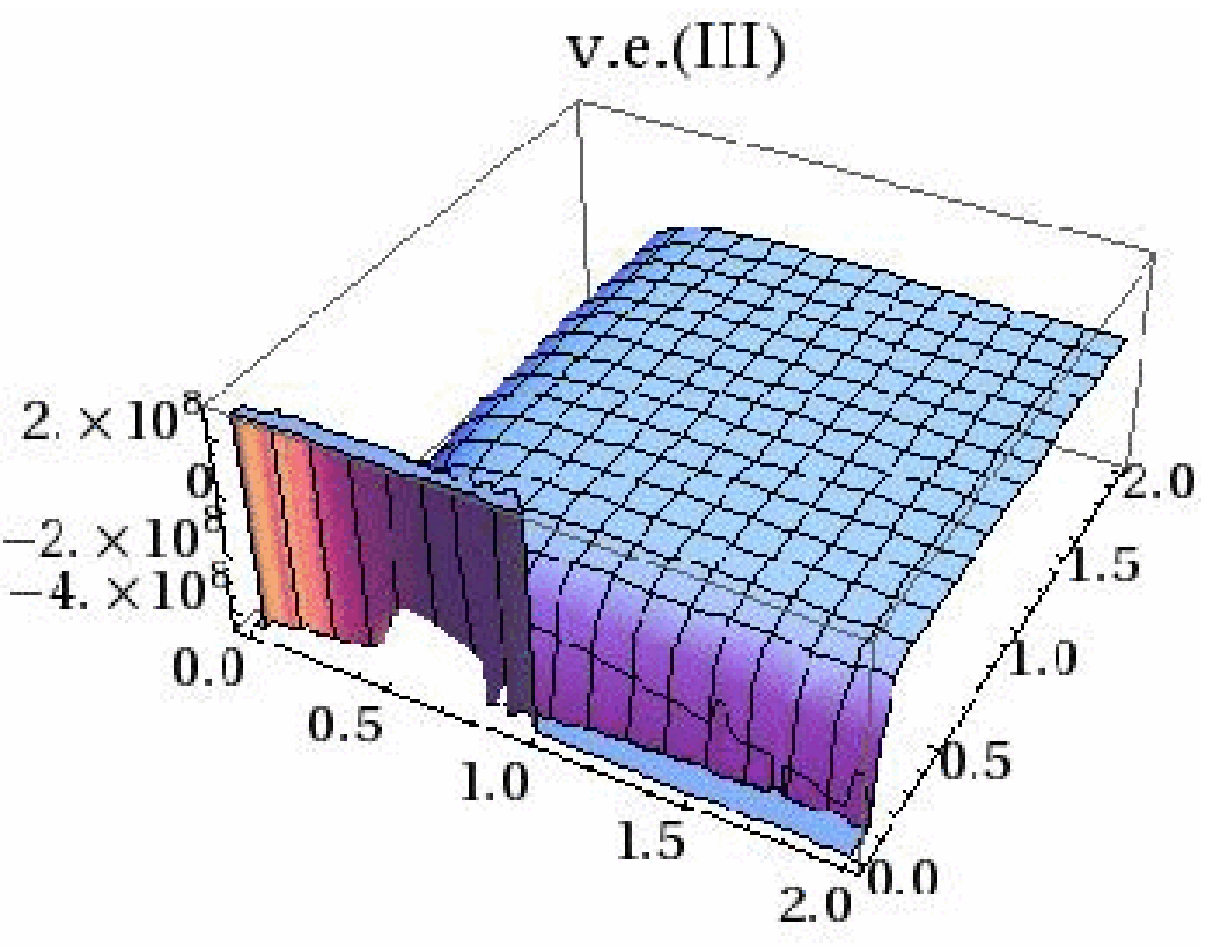}
\caption{ \label{Fig4} Plots of the contact interaction and of the vector-exchange contributions in the specialized planar configuration (minus sign).}
\end{figure}

\noindent In the equilateral configuration the four sides of the tetrahedron have the same length ($k_{1}=k_{2}=k_{3}=k_{4}$), therefore $x\equiv k_{\hat{12}}/k_{1}$ and $y\equiv k_{\hat{14}}/k_{1}$ can be chosen as variables for the plots. The plots of the isotropic functions of contact interaction and vector exchange contributions are provided in Fig.~4. The former ($c.i.$) shows a constant behaviour in this configuration, being independent of $k_{\hat{12}}$ and $k_{\hat{14}}$. The latter ($v.e.(I)$, $v.e.(II)$ and $v.e.(III)$) diverge as $k_{\hat{1i}}^{-3}$ ($i=1,2,3$ respectively for the three plots) in the limit of a flat tetrahedron, i.e. $(k_{\hat{1i}}/k_{1})\rightarrow 0$.

\noindent In the specialized planar configuration, the tetrahedron is flattened and, in addition to that, three of the six momentum variables are set equal to one another ($k_{1}=k_{3}=k_{\hat{14}}$); this leaves two independent variables, which can be $x\equiv k_{2}/k_{1}$ and $y\equiv k_{3}/k_{1}$. There is a double degeneracy in this configuration, due to the fact that the quadrangle can have internal angles larger than or smaller/equal to $\pi$, as we can see from the plus and minus signs in the expressions for $k_{\hat{12}}$ and $k_{\hat{13}}$ \cite{Chen:2009bc}
\bea\label{k12}
\frac{k_{\hat{12}}}{k_{1}}=\sqrt{1+\frac{x^2 y^2}{2}\pm\frac{xy}{2}\sqrt{(4-x^2)(4-y^2)}},\\\label{k13}
\frac{k_{\hat{13}}}{k_{1}}=\sqrt{x^2+y^2-\frac{x^2 y^2}{2}\mp\frac{xy}{2}\sqrt{(4-x^2)(4-y^2)}}.
\eea
The two cases are plotted in Figs.~5 and 6. Notice that divergences generally occur as $x,y \rightarrow 0$, as $x\rightarrow y$ and $(x,y)\rightarrow (2,2)$. 

\subsection{Features and level of anisotropy}

Statistical homogeneity and isotropy are considered characterizing features of the CMB fluctuations distribution, if one ignores the issues raised by the ``anomalous'' detections we presented in the introduction. \\
Homogeneity of the correlation functions equates translational invariance and hence total momentum conservation, as enforced by the delta functions appearing on the left-hand sides of Eqs.~(\ref{ps}) through (\ref{trisp}). This invariance property can then be pictured as the three momentum vectors forming a closed triangle for the bispectrum and the four momenta arranged in a tetrahedron for the trispectrum (see Fig.~7).\\
Statistical isotropy corresponds to invariance w.r.t. rotations in space of the momentum (for the power spectrum) and of the triangle or tetrahedron made up by the momenta, respectively for the bispectrum and the trispectrum. This symmetry can be broken, as it for example happens in the $SU(2)$ case, by assuming the existence of preferred spatial directions in the early universe that might be revealed in the CMB observations. When this happens, the correlation functions are expected to be sensitive to the spatial orientation of the wave number or of the momenta triangles and tetrahedrons w.r.t. these special directions. Analitically, the bispectrum and the trispectrum will depend on the angles among the vector bosons and the wave vectors (besides the angles among the gauge bosons themselves), as shown in the coefficients appearing in Eqs.~(\ref{fire}) and (\ref{fire1}). This implies that both the amplitude and the shape of bispectrum and trispectrum will be affected by these mutual spatial orientations. The modulation of the shapes by the directions that break statistical anisotropy was discussed with some examples both for the bispectrum and the trispectrum in our previous papers \cite{Bartolo:2009pa,Bartolo:2009kg}. These examples are here reported in Figs.~9 and 10. In Fig.~9 we show the plot of the vector contribution to the bispectrum of $\zeta$, properly normalized in the configuration
\bea\label{instance1}
\vec{N}_{3}=N_{A}(0,0,1)\\\label{instance2}
\vec{N}_{1}=\vec{N}_{2}=N_{A}(\sin\theta\cos\phi,\sin\theta\sin\phi,\cos\theta),
\eea
where, the $(x,y,z)$ coordinate frame is chosen to be $\hat{k}_{3}=\hat{x}$ and $\hat{k}_{1}=\hat{k}_{2}=\hat{z}$ and $\delta$ is the angle between $\vec{N}_{1,2}$ and $\hat{k}_{3}$.\\
In Fig.~10 we provide a similar plot for the trispectrum, but in a different configuration
\bea
\hat{N}_{2}\cdot\hat{k}_{i}=0\,(i=1,...4)\nonumber\\
\hat{N}_{1}\cdot\hat{k}_{1}=\cos\delta,\quad\quad\quad\hat{N}_{1}\cdot\hat{k}_{2}=0\nonumber\\
\hat{N}_{3}\cdot\hat{k}_{2}=\cos\theta,\quad\quad\quad\hat{N}_{3}\cdot\hat{k}_{1}=0.
\eea 
In both examples, it is assumed for simplicity that the $\vec{N}_{a}$ have the same magnitude $N_{A}$ for all $a=1,2,3$.\\

\begin{figure}\centering
 \includegraphics[width=0.37\textwidth]{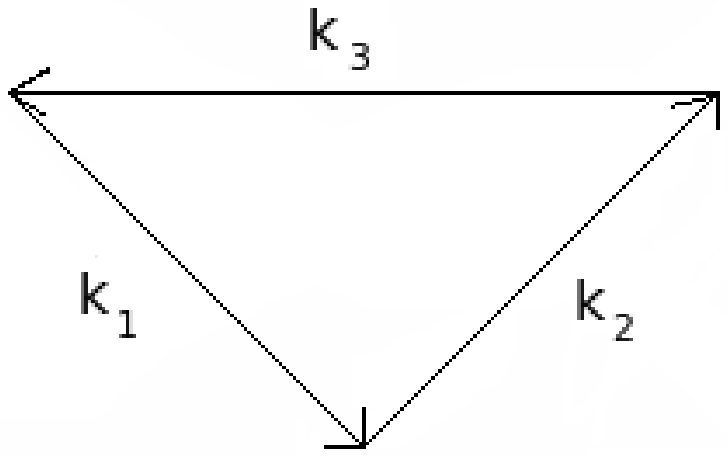}
\hspace{0.17\textwidth}
 \includegraphics[width=0.37\textwidth]{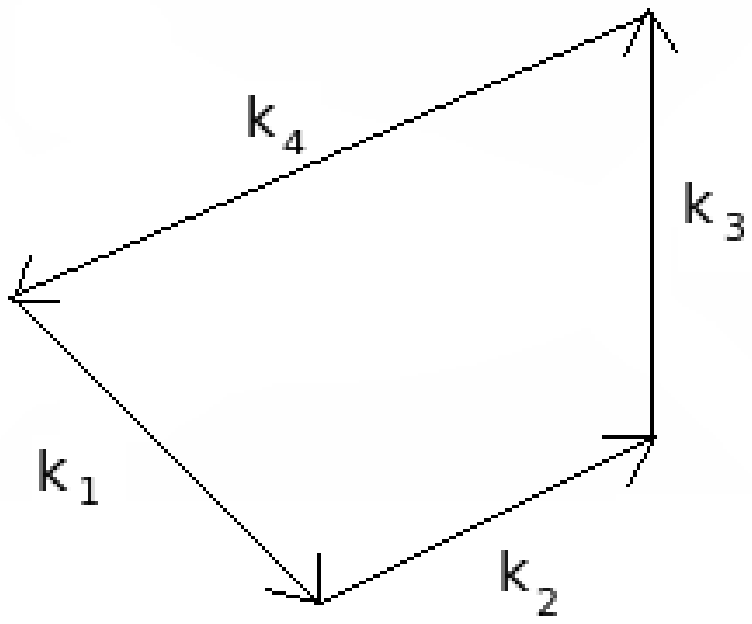}
\caption{ \label{Fig1} Representation of momentum conservation for the bispectrum (the three momenta form a closed triangle) and for the trispectrum (the momenta form a tetrahedron).}
\end{figure}

\noindent Another comment should be added concerning statistical anisotropy in the model. Notice that both the bispectrum and the trispectrum can be written as the sum of a purely isotropic and an anisotropic parts. The orders of magnitude of these two parts can, for instance, be read from Table 2 for the trispectrum: each one among $\tau_{NL}^{NA_{2}}$, $\tau_{NL}^{A_{1}}$ and $\tau_{NL}^{A_{2}}$ provide the order of magnitude of the level of both their isotropic and anisotropic contributions, which are therefore comparable; $\tau_{NL}^{NA_{1}}$ instead quantifies a purely anisotropic contribution which, as discussed in Sec.~5, can be comparable to the other three parts, if not the dominant one. A similar discussion applies to the bispectrum (see $f_{NL}^{A}$ and $f_{NL}^{NA}$ in Table 1). We can then conclude that for the three and for the four point function, there is room in the parameter space of the theory for the anisotropic contributions to be as large as, or even larger than, the isotropic ones.

\begin{figure}\centering
 \includegraphics[width=0.4\textwidth]{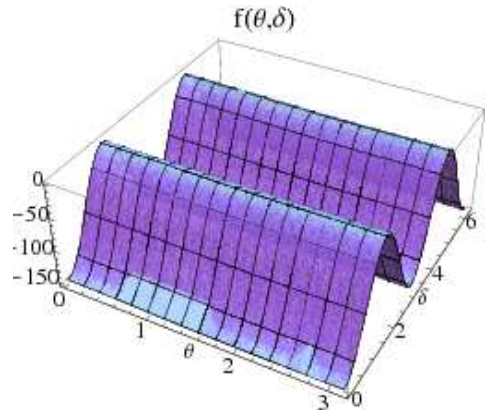}
\caption{ \label{Fig2}Plot of $f(\theta,\delta)\equiv {[(B_{\zeta}(\theta,\delta,x^{*},x_{2},x_{3})x_{2}^{2}x_{3}^{2}k_{1}^{6})/(g_{c}^{2}H^2 m^2 N_{A}^{4})]}$ evaluated at\\${(x^{*}=1,x_{2}=0.9,x_{3}=0.1)}$ in a sample angular configuration. See Appendix D of \cite{Bartolo:2009pa} for its complete analytic expression.}
\end{figure} 

\begin{figure}\centering
\includegraphics[width=0.4\textwidth]{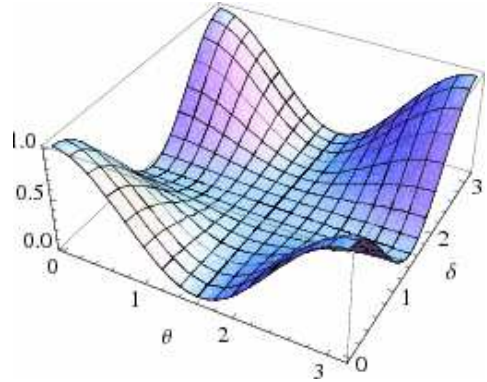}
\caption{ \label{Fig5} Plot of the anisotropic part of the trispectrum from the contribution due to vector-exchange diagrams in a sample angular configuration. See Sec.~8 of \cite{Bartolo:2009kg} for its analytic expression.}
\end{figure}

\section{Conclusions}

Motivated by the interest in models that combine non-Gaussianity and statistical anisotropy predictions for the CMB fluctuations, we have considered models of inflation where primordial vector fields effectively participate in the production of the curvature perturbations $\zeta$. More specifically, we have reviewed the computation of the correlation functions up to fourth order, considering an $SU(2)$ vector multiplet. The $\delta$N formalism was employed to express $\zeta$ in terms of the quantum fluctuations of all the primordial fields. The Schwinger-Keldysh formula was also used in evaluating the correlators.\\

\noindent The correlation functions result as the sum of scalar and vector contributions. The latter are of two kinds, ``Abelian'' (i.e. arising from the zeroth order terms in the Schwinger-Keldysh expansion) and ``non-Abelian'' (i.e. originating from the self-interactions of the vector fields). The bispectrum and the trispectrum final results are presented as a sum of products of isotropic functions of the momenta, $F_{n}$ and $G_{n}$ in Eqs.~(\ref{fire}) and (\ref{fire1}), multiplied by anisotropy coefficients, $I_{n}$ and $L_{n}$ in (\ref{fire}) and (\ref{fire1}), which depend on the angles between the (gauge and wave) vectors. \\

\noindent The amplitude of non-Gaussianity has been presented through the parameters $f_{NL}$ and $\tau_{NL}$; in particular we have show the dependence of these functions from the non-angular parameters of the theory. We have provided the comparisons among the different (scalar versus vector, Abelian versus non-Abelian) contributions to $f_{NL}$ and $\tau_{NL}$, noticing that any one of them can be the dominant contribution depending on the selected region of parameter space. In particular, we have stressed how the anisotropic contributions to the bispectrum and the trispectrum can overcome the isotropic parts. An interesting feature of these models is that the bispectrum and the trispectrum depend on the same set of parameters and their amplitudes are therefore strictly related to one another. \\

\noindent We have presented the shapes of both the bispectrum and the trispectrum. The isotropic functions appearing in their final expressions had been analyzed separately from their anisotropy coefficients. The bispectrum isotropic functions had been found to preferably show a local shape. The trispectrum ones had been plotted selecting equilateral and specialized planar configurations. The full expressions (i.e. complete of their anisotropy coefficients) of bispectrum and trispectrum have been presented in specific momenta configuration, in order to provide a hint of the modulation of shapes and amplitudes operated by anisotropy.\\

\noindent We have reviewed old and recent vector field models, indicating both their limits and achievements. We would like to stress that, in our view, the most promising features of these models consists in the possibility of providing both non-Gaussianity and statistical anisotropy predictions that are related to one another because of the fact that they share the same underlying theory. This might, at some point in the future, become a great advantage: measurements of non-Gaussianity could be used to constrain statistical anisotropy or viceversa.

\section*{Acknowledgments}

This research has been partially supported by ASI contract I/016/07/0 ``COFIS''. A.R. acknowledges support by the EU Marie Curie Network UniverseNet (HPRNCT2006035863).




\end{document}